\documentclass[twocolumn,english,prl,groupedaddress]{revtex4}
\usepackage[T1]{fontenc}
\usepackage[latin9]{inputenc}
\usepackage{color}
\usepackage{verbatim}
\usepackage{textcomp}
\usepackage{mathrsfs}
\usepackage{mathtools}
\usepackage{amsmath}
\usepackage{amssymb}
\usepackage{graphicx}
\usepackage{dblfloatfix}
\usepackage{braket}
\usepackage{overpic}
\usepackage{bm}

\makeatletter

\pdfpageheight\paperheight
\pdfpagewidth\paperwidth

\providecommand{\tabularnewline}{\\}

\@ifundefined{textcolor}{}
{
 \definecolor{BLACK}{gray}{0}
 \definecolor{WHITE}{gray}{1}
 \definecolor{RED}{rgb}{1,0,0}
 \definecolor{GREEN}{rgb}{0,1,0}
 \definecolor{BLUE}{rgb}{0,0,1}
 \definecolor{CYAN}{cmyk}{1,0,0,0}
 \definecolor{MAGENTA}{cmyk}{0,1,0,0}
 \definecolor{YELLOW}{cmyk}{0,0,1,0}
}

\newcommand{\ketbra}[3][]{%
  \def\@tempa{#1}%
  \ifx\@tempa\@empty\relax%
    \mathinner{\lvert#2\rangle\langle #3\rvert}%
  \else%
    \mathinner{\lvert#2\rangle\langle #3\rvert}_{#1}%
  \fi}

\RequirePackage[colorlinks=true,linkcolor=blue]{hyperref}
\usepackage{dsfont}

\AtBeginDocument{
  
}

\makeatother

\usepackage{babel}

\begin{document}

%% Title
\title{Solid-state magnetic traps and lattices}

%% Authors
\author{J. Kn\"orzer,$^{1,*}$ M. J. A. Schuetz,$^{2,*}$ G. Giedke,$^{3,4}$
H. Huebl,$^{5,6,7}$ M. Weiler,$^{5,6}$ M. D. Lukin,$^{2}$ and J. I. Cirac$^{1}$ }

%% Affiliations
\affiliation{$^{1}$Max-Planck-Institut f\"ur Quantenoptik, Hans-Kopfermann-Str.
1, 85748 Garching, Germany}
\affiliation{$^{2}$Physics Department, Harvard University, Cambridge, MA 02318,
USA}
\affiliation{$^{3}$Donostia International Physics Center, Paseo Manuel de Lardizabal
4, E-20018 San Sebasti\'{a}n, Spain}
\affiliation{$^{4}$Ikerbasque Foundation for Science, Maria Diaz de Haro 3, E-48013
Bilbao, Spain}
\affiliation{$^{5}$Walther-Mei{\ss}ner-Institut, Walther-Mei{\ss}ner-Str. 8, 85748 Garching, Germany}
\affiliation{$^{6}$Physik-Department, Technische Universit\"at M\"unchen, 85748 Garching, Germany}
\affiliation{$^{7}$Nanosystems Initiative Munich (NIM), Schellingstra{\ss}e 4, 80799 M\"unchen, Germany}

\date{\today}

%% %%%%%%%%%%%%%%%%%%%%%%%%%%%%%%%%%%%%%%%%%%%%%%%%%%%%%%%%%%%%%%%%%%%%%
%% Abstract
\begin{abstract}
We propose and analyze magnetic traps and lattices for electrons in
semiconductors. 
We provide a general theoretical framework and show that thermally
stable traps can be generated by magnetically driving the particle's
internal spin transition, akin to optical dipole traps for ultra-cold
atoms. 
Next we discuss in detail periodic arrays of magnetic traps,
i.e.~\textit{magnetic lattices}, as a platform for quantum simulation of
exotic Hubbard models, with lattice parameters that can be tuned in real
time. 
Our scheme can be readily implemented in state-of-the-art experiments,
as we particularize for two specific setups, one based on a
superconducting circuit and another one based on surface acoustic waves.
\end{abstract}

\maketitle

%% %%%%%%%%%%%%%%%%%%%%%%%%%%%%%%%%%%%%%%%%%%%%%%%%%%%%%%%%%%%%%%%%%%%%%
%% Section 1:
%% INTRODUCTION
\section{Introduction }

%% Executive summary
The advent of cold atoms trapped in optically defined potential landscapes has
enabled experimental breakthroughs in various discplines ranging from
condensed-matter physics to quantum information processing \cite{lewenstein07,bloch12}.
Especially, thanks to largely tunable system parameters and the possibility to
mimic and gain understanding of complex solid-state systems, ultra-cold
atomic gases have become a rich playground and valuable tool to explore
novel quantum many-body physics \cite{bloch08}.
On a complementary route towards controllabe quantum matter and a fully
fledged quantum simulator, solid-state platforms allow to pursue the
same goals in a very different physical context,
both bearing challenges such as to overcome
impurity-induced disorder in semiconductor systems \cite{barthelemy13},
but also offering the potential to benefit from long-range inter-particle
interactions, access to a wide variety of quasiparticles and, in principle,
means to build scalable on-chip architectures for quantum information processing.
To this end, different kinds of quasiparticle traps
in semiconductor nanostructures have been proposed and realized
\cite{lee01,hanson07,alloing13,schuetz10,rocke97,zimmermann99,ford17}.
Likewise, in the realm of atomic \cite{folman02,keil16}
and molecular \cite{andre06,hou17} systems, mesoscopic on-chip platforms
have been tailored to miniaturize experiments with ultracold quantum matter.
Apart from more established solid-state platforms like, e.g., quantum-dot based
architectures \cite{hensgens17}, it has recently been proposed \cite{schuetz17}
to employ surface acoustic waves (SAWs) to trap and control semiconductor
quasiparticles such as electrons in intrinsically scalable and tunable \textit{acoustic lattices}.
The latter operate at elevated energy scales with typical
lattice spacings $a\gtrsim100~$nm and recoil energies $E_\text{R}/k_\text{B}\sim(0.1-1)~K$
(where $E_\text{R}=h^2/(8ma^2)$ with an effective particle mass $m$ which is
typically of the order of the electron rest mass)
as compared with optical lattices where typically $E_\text{R}/k_\text{B}\sim10^{-7}K$
\cite{romero-isart13}.
Inspired by these results and recent advances in the rapidly evolving
field of nanomagnetism \cite{tejada17,salasyuk17}, i.e., the generation
and control of (high-frequency) magnetic fields on the nanoscale,
the present work aims to bring the favourable scaling properties and
flexibility of optical lattices to the solid-state domain.

In contrast to electrically defined confinement potentials for charged
particles in quantum wells,
the spin degree of freedom (DOF) can be addressed with magnetic field
gradients in
order to trap and control particles in semiconductor nanostructures;
note that this is in close analogy to the working principle of
optical dipole traps where the induced AC Stark shift of
the atomic levels gives rise to a dipole potential for the atom \cite{grimm00}.
In previous theoretical proposals \cite{redlinski05,berciu05} and
experimental demonstrations \cite{christianen98,murayama06},
magnetic traps for charge carriers in low-dimensional quantum wells were
induced by a spatially inhomogeneous \textit{giant Zeeman splitting} in
dilute magnetic semiconductors (DMS) \cite{furdyna88}, which feature
extremely large $g$-factors $\sim 10^2$.
In particular, microscale magnets \cite{halm08} and current loops \cite{chen08} as well
as superconducting (SC) vortex lattices \cite{berciu05} have been considered in this
context.
So far, however, none of these previous results have yet been tailored
to scalable architectures and, moreover, only static traps with limited
tunability of system parameters have been taken into account.
In this work,
we take a significant next step towards tunable and scalable
magnetic lattices and develop a general theoretical framework fit to
describe the latter.
We show that a non-standard form of the Hubbard model with
two independently tunable hopping parameters can readily be implemented.
Ultimately,
two alternative implementations of the developed model will be discussed
in detail, one based on SAWs and the other based on magnetic field gradients
generated by SC nanowires, both operated
in yet unexplored parameter regimes and with highly
favourable tunability and scalability properties.

The basic scheme is depicted in Fig.~\ref{fig:setup}.
We consider electrons with two internal (spin) states $\ket{\uparrow}$ and
$\ket{\downarrow}$ which
are confined to a conventional low-dimensional quantum well or a
purely two-dimensional material, e.g., from the group of transition-metal
dichalcogenides (TMDs),
and subject to a spatially inhomogeneous magnetic
driving field.
Due to the thereby induced AC Stark shift acting on the internal energy
levels, the electrons feel an effective state-dependent
potential which is periodic along one axis (in the one-dimensional setup
we consider here), as illustrated in Fig.~\ref{fig:setup}.
As a result, the electrons are attracted to a regular lattice of
antiferromagnetic character, since the two internal states are found to
be trapped at nodes or antinodes of the magnetic field distribution,
respectively, cf.~Fig.~\ref{fig:setup}(b).
For simplicity, we consider only one-dimensional systems, but all results
can readily be generalized to two dimensions.

This paper is organized as follows.
In Sec.~\ref{sec:theory}, we first introduce the theoretical framework to describe
magnetic trapping potentials for electrons confined to a two-dimensional
electron gas (2DEG).
All requirements for the validity of the theoretical treatment and
relevant approximations are discussed in Sec.~\ref{ssec:single-particle},
followed by an investigation of hopping and interactions in magnetic lattices
[see Sec.~\ref{ssec:many-body}] and a detailed description
of possible implementations in Sec.~\ref{sec:implementations}.
Finally, we will provide case studies for both implementations with
realistic parameters in Sec.~\ref{sec:case-study}.

%% %%%%%%%%%%%%%%%%%%%%%%%%%%%%%%%%%%%%%%%%%%%%%%%%%%%%%%%%%%%%%%%%%%%%%

%~ %% %%%%%%%%%%%%%%%%%%%%%%%%%%%%%%%%%%%%%%%%%%%%%%%%%%%%%%%%%%%%%%%%%%%%%
%~ %% Section 2:
%~ %% THEORETICAL FRAMEWORK
\section{General theoretical Framework \label{sec:theory}}

\subsection{Single-particle physics in magnetic traps \label{ssec:single-particle}}
\textit{Single-particle physics}.\textemdash
We consider an electron confined to a 2DEG with effective mass $m$ and the two internal states
$\ket{\uparrow}$ and $\ket{\downarrow}$
exposed to an external magnetic field,
$\mathbf{B}(\mathbf{r},t)=\mathbf{B}_{\perp}(\mathbf{r},\omega t) + \mathbf{B}_{||}$.
The spatially homogeneous, static (in-plane) part of the field,
$\mathbf{B}_{||} = B_0 \hat{\mathbf{z}}$, gives rise to a
Zeeman splitting, $\hbar \omega_0 = g_\text{s} \mu_\text{B} B_0$, and the
inhomogeneous (\textit{time-dependent} or \textit{time-independent})
(out-of-plane) field component,
$\mathbf{B}_{\perp}(\mathbf{r},\omega t) = B_1 \Lambda(\mathbf{r}) \cos(\omega t) \hat{\mathbf{x}}$,
drives spin transitions with frequency $\omega$.
The corresponding Hamiltonian can be written as (here and in the following, we adopt the convention that $\hbar = 1$)
%~ %% %%%%%%%%%%%%%%%%%%%%%%%% EQUATION 1 %%%%%%%%%%%%%%%%%%%%%%%%%%%%%%%%%%%
\begin{equation}\label{eq:model}
H = \frac{\hat p^2}{2m} + h(\hat z) = \frac{\hat p^2}{2m} + \frac{\omega_0}{2} \sigma^z + \frac{\Omega(\hat z)}{2} \cos(\omega t) \sigma^x,
\end{equation}
%~ %% %%%%%%%%%%%%%%%%%%%%%%%% EQUATION 1 %%%%%%%%%%%%%%%%%%%%%%%%%%%%%%%%%%%
where $\hat z$, $\hat p$, $\sigma^x = \ket{\uparrow}\bra{\downarrow} + \ket{\downarrow}\bra{\uparrow}$,
$\sigma^z = \ket{\uparrow}\bra{\uparrow} - \ket{\downarrow}\bra{\downarrow}$ denote the particle's position,
momentum and Pauli spin operators, respectively.
The inhomogeneous Rabi frequency is denoted by
$\Omega(\hat z) = \Omega_0 \Lambda(\hat z)$ with $\Omega_0 = \gamma B_1$,
where $\gamma = g_\text{s} \mu_\text{B}$ is the gyromagnetic
ratio of the electron.
We assume $\Lambda(\hat z) = \cos(k \hat z)$ in the following, where $k$
denotes the wavevector, but more general periodic functions can be
considered.
While the universality of this model will become more
apparent later, especially when we consider different implementations in
Sec.~\ref{sec:implementations}, we may already distinguish between two
physically dissimilar cases both captured by Eq.~(\ref{eq:model}):
(\textit{i}) \textit{static} traps ($\omega = 0$) are time-independent
and (\textit{ii}) \textit{dynamic} traps ($\omega > 0$) are explicitly
time-dependent realizations of the model.
Due to their intrinsic flexibility and \textit{in-situ} tunability of
system parameters, we put the main focus on \textit{dynamic} magnetic
traps, i.e., $\omega>0$.
%% %%%%%%%%%%%%%%%%%%%%%%%% FIGURE 1 %%%%%%%%%%%%%%%%%%%%%%%%%%%%%%%%%%%
\begin{figure}[t!]
\centering
\includegraphics[width=0.49\textwidth]{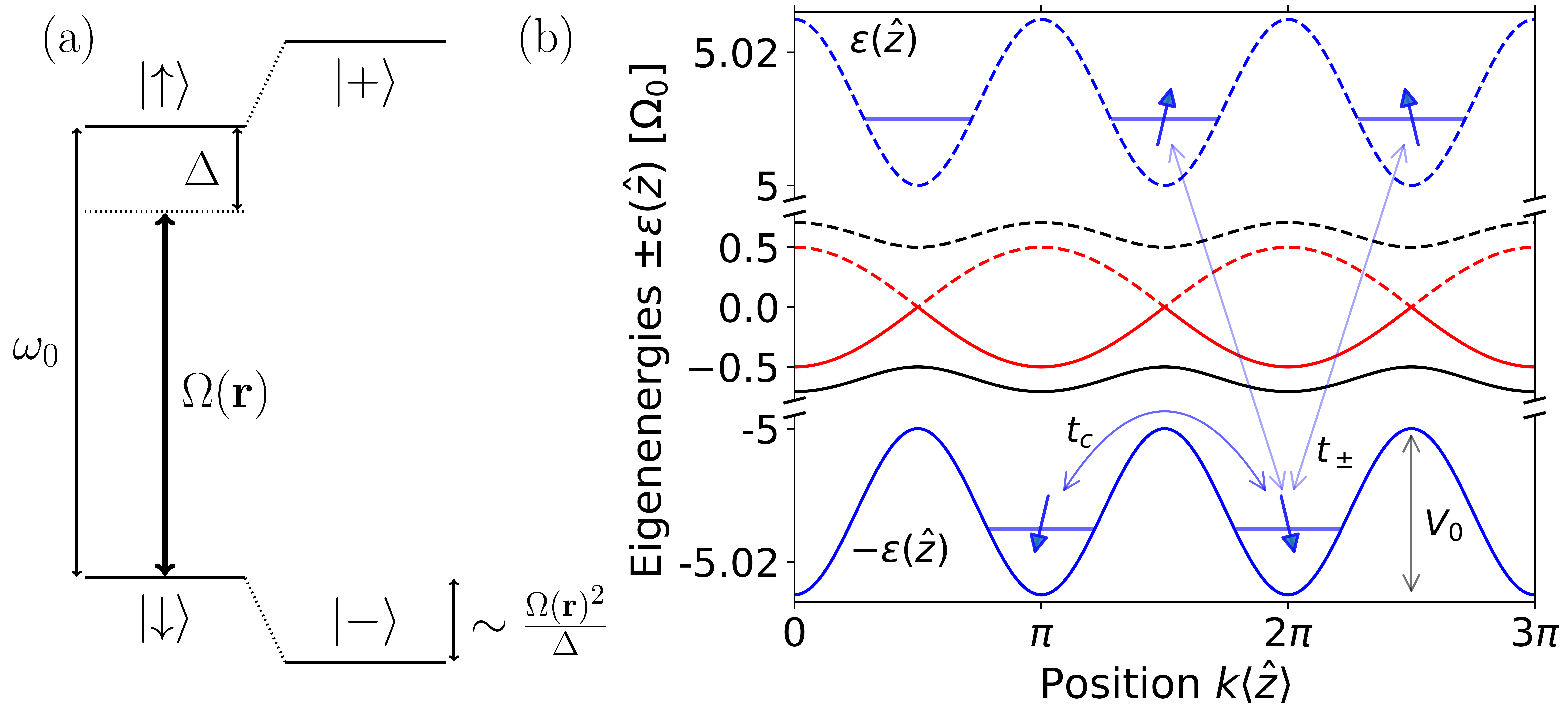}
\caption{\label{fig:setup}(color online). Schematic illustration of the
trapping scheme and magnetic lattice.
(a) At each point, the two-level spin systems experience an AC Stark shift which
defines an effective (state-dependent) potential landscape for the electrons.
(b) The local eigenenergies $\pm\varepsilon(\hat z)$ of the two
spin components $\ket{+}_{\theta(\hat z)}$ and $\ket{-}_{\theta(\hat z)}$ are shifted with
respect to each other.
The energies $+\varepsilon(\hat z)$ (dashed) and $-\varepsilon(\hat z)$ (solid) are
shown for $\Delta/\Omega_0 = 10$ (blue), $\Delta/\Omega_0 = 1$ (black)
and $\Delta/\Omega_0 = 0$ (red) in units of $\Omega_0$.
The hopping matrix elements [see Sec.~\ref{ssec:many-body}] $t_c$ and $t_{\pm}$
denote next-nearest neighbour spin-conserved and nearest-neighbour
spin-flip assisted tunneling, respectively.
$V_0$ denotes the trap depth.}
\end{figure}
%% %%%%%%%%%%%%%%%%%%%%%%%% FIGURE 1 %%%%%%%%%%%%%%%%%%%%%%%%%%%%%%%%%%%

Within a co-rotating frame and rotating-wave approximation (RWA) for $|\Delta| = |\omega_0 - \omega|
\ll \omega_0 + \omega$ and $\Omega_0 \ll \omega$, the time-independent internal model
$h_\text{RWA}(\hat z) = [\Delta/2] \sigma^z + [\Omega(\hat z)/2] \sigma^x$ can be diagonalized
exactly which yields the local eigenenergies
$\pm \varepsilon(\hat z)$ with $\varepsilon(\hat z)=\frac{1}{2} \sqrt{\Omega^2(\hat z)+\Delta^2}$
and position-dependent eigenstates,
\begin{equation}\label{eq:exact-eigenstates}
\begin{gathered}
\ket{+}_{\theta(\hat z)} = \cos \frac{\theta(\hat z)}{2} \ket{\uparrow} + \sin \frac{\theta(\hat z)}{2} \ket{\downarrow}, \\
\ket{-}_{\theta(\hat z)} = -\sin \frac{\theta(\hat z)}{2} \ket{\uparrow} + \cos \frac{\theta(\hat z)}{2} \ket{\downarrow}, \nonumber \\
\end{gathered}
\end{equation}
where $\theta(\hat z) = \arcsin [ \frac{\Omega(\hat z)}{\sqrt{\Omega^2(\hat z)+\Delta^2}}  ]$.
The trap depth of the effective potentials $\pm\varepsilon(\hat z)$, which is
given by the difference $|\underset{z}{\mathrm{max}}\ \varepsilon(\hat z)-\underset{z}{\mathrm{min}}\ \varepsilon(\hat z)|$,
depends only on $\Omega_0$ and $\Delta$ and will be denoted by $V_0$ in the
following [see Fig.~\ref{fig:setup}].
In the limit $\Omega_0 \ll |\Delta|$, the standard result from
second-order perturbation theory, $\varepsilon(\hat z) \approx |\Delta|/2 + \Omega_0^2 \Lambda^2(\hat z)/(4|\Delta|)$,
can be recovered.
Note that the periodic modulation of the internal energy levels
$\ket{\pm}_{\theta(\hat z)}$ amounts to a
state-dependent potential for the motional DOF
such that the states are trapped at nodes
and antinodes of the driving field, respectively.
As a consequence, magnetic trapping potentials for the two spin
components are shifted with respect to one another, as illustrated in Fig.~\ref{fig:setup}(b).
In fact, this result is reminiscent of state-dependent optical lattices
which can be enriched by laser-assisted tunneling between internal atomic
states \cite{jaksch98,ruostekoski02}, whereby gauge fields for ultracold
atoms can be generated \cite{jaksch03,dalibard11,goldman14,goldman16}.

Note that, in the realm of the RWA introduced before, the Rabi frequency
$\Omega_0$ is limited to relatively small values, as compared to other
relevant energy scales.
This limitation can be overcome, to some extent, by deriving an effective
Floquet Hamiltonian without RWA, see Appendix \ref{sec:beyond-rwa} for details.

Until now, we have not explicitly taken into account the presence of the
kinetic term, $\hat p^2/(2m)$, in Eq.~(\ref{eq:model}).
Its presence induces a coupling between the local spin eigenstates
$\ket{\pm}_{\theta(\hat z)}$ and, as a consequence, undesired spin flips
may result in particle loss from the trap \cite{sukumar97}.
In order to quantify this effect, it is instructive to introduce a
unitary transformation $U(\hat z)$ which diagonalizes $h_\text{RWA}(\hat z)$
at each point, such that $\ket{+}_{\theta(\hat z)} = U(\hat z) \ket{\uparrow}$
($\ket{-}_{\theta(\hat z)} = U(\hat z) \ket{\downarrow}$).
The thereby transformed Hamiltonian,
$\tilde H = U^\dagger [\frac{p^2}{2m} + h_\text{RWA}(\hat z)] U = \hat p^2/2m + \tilde h(\hat z) + \Delta T$,
contains the kinetic term from Eq.~(\ref{eq:model}), the diagonal
(in the local eigenbasis spanned by $\ket{ + }_{\theta(\hat z)}$ and $\ket{ - }_{\theta(\hat z)}$)
spin Hamiltonian $\tilde h = U^\dagger h_\text{RWA} U$ and an additional
term $\Delta T$, which stems from the transformation of the kinetic term,
see Appendix \ref{sec:non-adiabatic-transitions} for details.
If the latter contributes only a small correction to the system's
characteristic energy scale set by the motional quantum $\omega_\text{HO}$,
the internal spin DOF follows adiabatically the local direction of the
magnetic field and the contribution from $\Delta T$ can be safely neglected.
For this \textit{adiabatic approximation}
(also refered to as Born-Oppenheimer approximation) to hold,
the local eigenstates of the two-level system spanned by $\ket{+}_{\theta(\hat z)}$
and $\ket{-}_{\theta(\hat z)}$ must be sufficiently separated in energy.
If this energy gap by far exceeds $\omega_\text{HO}$,
i.e.~$\chi := \omega_\text{HO} / |\Delta| \ll 1$,
spin-flip processes are typically negligible \cite{sukumar97}.

%% Requirements
\textit{Requirements}.\textemdash
Following the line of arguments outlined above, we have implicitly made
a few assumptions about the system parameters which we are going to
summarize in the following:
(\textit{i}) We have assumed idealized two-level spin systems
with well-resolved energy levels and thus a
relatively small intrinsic linewidth $\Gamma \ll |\Delta|$.
(\textit{ii}) We require a weak electron-phonon coupling, i.e.,
the spontaneous phonon emission rate $\gamma$ which quantifies
motional damping of the electron must be small compared to all other
characteristic system's time scales; explicitly, we demand that it should
be smaller than the motional transition frequencies, i.e., $\gamma \ll \omega_\mathrm{HO}$.
(\textit{iii}) In order to obtain thermally robust traps and minimize particle loss
from the trap, we need thermal energies $k_\text{B} T \ll V_0$
(where $k_\text{B}$ denotes the Boltzmann constant).
Typically, in case ground-state cooling is desired, this requirement is
replaced by the stronger condition $k_\text{B} T \ll \omega_\text{HO}$.
(With at least one bound state, $n_b = V_0 / \omega_\text{HO} \geq 1$,
supported by the trap, the latter condition is more restrictive.)
(\textit{iv}) The magnetic trap depth $V_0$ is either much smaller than
$\Omega_0$, i.e.~$V_0 = \Omega_0^2/(4|\Delta|)$ in the perturbative regime
$\Omega_0 \ll |\Delta|$, or approaches $V_0 \rightarrow \Omega_0 / 2$ in
the opposite limit $|\Delta|/\Omega_0 \rightarrow 0$;
however, in both cases $V_0$ is limited from above by $\Omega_0/2$.
In terms of other relevant physical parameters contained in
$\Omega_0=\gamma B_1$, this means that strong
magnetic radio-frequency (RF) fields $\sim B_1$ and large g-factors are favourable.
(\textit{v}) The Rabi frequency $\Omega_0$, in turn, is typically much smaller than the
driving frequency within the RWA, $\Omega_0 \ll \omega$, but this condition
can be relaxed as mentioned earlier.
However, for too large $\Omega_0$, even the high-frequency expansion of
the Floquet Hamiltonian fails to converge.
For our purposes, we therefore demand $\Omega_0 < \omega$.
(\textit{vi}) Finally,
introducing the small number $\varepsilon_\text{ad} = V_0 / \omega \lesssim 0.5$, the
adiabaticity condition $\chi \ll 1$ can be rewritten as
$\omega \ll n_b |\Delta| / \varepsilon_\text{ad}$.
However, this condition may be relaxed at the cost of higher loss rates.
The Majorana loss rate $\Gamma_\text{loss}$, compared to the natural frequency
scale $\omega_\text{HO}$ of the trap, can be estimated as
$\eta := \Gamma_\text{loss} / \omega_\text{HO} \approx 2\pi \exp \left ( -4/\chi \right )$
\cite{sukumar97}
(compare also Ref.~\cite{burrows17} for a related
description of non-adiabatic spin-flips in radio-frequency dressed
magnetic traps for cold atoms);
deep in the adiabatic regime with $\chi = 0.1$, spin-flip
losses are negligible as $\eta \sim 10^{-17}$, but even for moderate
values $\chi = 0.5$ ($\chi = 1$), the loss rates are relatively small
with $\eta \approx 2 \cdot 10^{-3}$ ($\eta \approx 1.2 \cdot 10^{-1}$).
Hence, the adiabaticity condition may be relaxed in order to obtain
well-performing traps.
Putting these findings together results in a concise list of necessary
requirements and,
in general, without resorting to the RWA or the perturbative regime
where $\Omega_0 \ll |\Delta|$, we find:
%~ %% %%%%%%%%%%%%%%%%%%%%%%%% EQUATION 2 %%%%%%%%%%%%%%%%%%%%%%%%%%%%%%%%%%%
\begin{equation}\label{eq:requirements}
\gamma, k_\text{B} T \ll \omega_\text{HO} \lesssim V_0 \lesssim \Omega_0/2 \lesssim \omega/2.
\end{equation}
%~ %% %%%%%%%%%%%%%%%%%%%%%%%% EQUATION 2 %%%%%%%%%%%%%%%%%%%%%%%%%%%%%%%%%%%
In order to obtain reliable magnetic traps, both implementations
discussed in Sec.~\ref{sec:implementations} need to be operated in a
parameter regime where Eq.~(\ref{eq:requirements}) is fulfilled and
$\eta$ is sufficiently small.

\subsection{Engineering of Hubbard models \label{ssec:many-body}}
\textit{Towards many-body physics}.\textemdash
Based on the theoretical framework fit to describe single traps as worked
out above, the following paragraphs are dedicated to the study of
Fermi-Hubbard physics in magnetic lattices, i.e., periodic arrays of magnetic traps.
Explicitly, we show that spin-dependent forms of the Hubbard model with
independently tunable hopping parameters $t_c$ and $t_\pm$
can be realized with the aid of additional driving fields
[see Appendix \ref{app:hubbard} for more details]
in the fashion of zigzag optical lattices for cold atoms \cite{greschner13,dhar13}.
Here and in the following, $t_c$ denotes spin-conserved
next-nearest neighbour coherent tunneling processes and $t_\pm$ describes spin flip-assisted
tunneling between adjacent lattice sites, cf.~Fig.~\ref{fig:setup}(b).
Another genuine prospect is the operation in a low-temperature, strong-interaction
regime (at dilution-fridge temperatures $T \approx (10-100)~$mK) where
the thermal energy is much smaller than the hopping parameters
$t_c, \ t_\pm$ which, in turn, are small compared to the on-site
interaction strength $U$, i.e., $k_\text{B} T \ll t_c, t_\pm < U$.

As a starting point, we consider the single-particle Hamiltonian $\tilde H$
within the adiabatic approximation [see Sec.~\ref{ssec:single-particle}]
which can be written as
%~ %% %%%%%%%%%%%%%%%%%%%%%%%% EQUATION 3 %%%%%%%%%%%%%%%%%%%%%%%%%%%%%%%%%%%
\begin{equation}\label{eq:adiabatic-model}
\tilde H \approx \frac{\hat p^2}{2m} + \tilde h(\hat z) = \frac{\hat p^2}{2m} + \varepsilon (\hat z) {\tilde \sigma}^z,
\end{equation}
%~ %% %%%%%%%%%%%%%%%%%%%%%%%% EQUATION 3 %%%%%%%%%%%%%%%%%%%%%%%%%%%%%%%%%%%
with ${\tilde \sigma}^z = |+\rangle\langle +| - |-\rangle\langle -|$.
In a next step, we now consider an ensemble of electrons in a magnetic lattice.
At sufficiently low temperatures ($k_\text{B} T \ll \omega_\text{HO}$)
such that the electrons are confined to the lowest Bloch band,
we find that the system is characterized by a Fermi-Hubbard model of the form \cite{byrnes07}
%~ %% %%%%%%%%%%%%%%%%%%%%%%%% EQUATION 4 %%%%%%%%%%%%%%%%%%%%%%%%%%%%%%%%%%%
\begin{eqnarray}\label{eq:hubbard}
H_{\text{FH}} & = & -t_c\sum_{\langle\langle i,j\rangle\rangle,s}(c_{is}^{\dagger}c_{js}+\text{h.c.})-
\varepsilon\sum_{is} (-1)^i n_{is} \nonumber \\
 &  & + \sum_{i} \mu_i n_i + \sum_{s,s^\prime} \sum_{ijkl} U_{ijkl} c_{is^\prime}^{\dagger}c_{js}^{\dagger}c_{ls}c_{ks^\prime},
\end{eqnarray}
%~ %% %%%%%%%%%%%%%%%%%%%%%%%% EQUATION 4 %%%%%%%%%%%%%%%%%%%%%%%%%%%%%%%%%%%
where the fermionic operator $c_{is}^{(\dagger)}$ annihilates (creates)
an electron with spin $s = +,-$
at lattice site $i$, $n_{is} = c_{is}^{\dagger}c_{is}$ and $n_i = n_{i+} + n_{i-}$
are the spin-resolved and total occupation numbers, respectively.
The summation over $\langle\langle \cdot, \cdot \rangle\rangle$ is performed for
next-nearest neighbours (accordingly, $\langle \cdot, \cdot \rangle$
in Eq.~\eqref{eq:hubbard2} denotes a summation over neighbouring sites).
$U_{ijkl}=\int \mathrm{d}z \mathrm{d} z^\prime w_i^*(z^\prime) w_j^*(z)
U_\text{C}(z,z^\prime)w_k(z) w_l(z^\prime)$
quantifies the inter-particle interaction strength
($U = U_{iiii}$ denotes the on-site interaction strength),
where $w_i$ is a Wannier basis function which is typically strongly
localized around the respective lattice $i$.
Typically, it is inversely proportional to the lattice constant $a$,
depends on the dielectric constant $\epsilon$ of the substrate
and can be reduced with the aid of an additional metallic screening layer
positioned at a distance $d_\text{scr}$ from the 2DEG.
The screened Coulomb interaction can be written as $U_\text{C}=e^2 f_s(z,z^\prime)/(4\pi\epsilon|z-z^\prime|)$,
where $f_s=1-|z-z^\prime|/\sqrt{(z-z^\prime)^2+4d_\text{scr}^2}$ incorporates
screening \cite{byrnes07,footnote2}.
In Eq.~\eqref{eq:hubbard} the spin-dependent energy offset $\sim \varepsilon$
[see Fig.~\ref{fig:setup}] incorporates the remnant of the Zeeman splitting
(in the rotating frame) and the AC Stark shifts.
Moreover, the site-dependent chemical potential $\mu_i$ can take disorder
effects into account \cite{schuetz17}.
In the tight-binding limit where the potential is sufficiently deep, i.e.,
$E_\text{R} \ll V_0$ (with the recoil energy $E_\text{R} = k^2 / 2m$),
the hopping parameter is approximately given by
$t_c/E_R\approx(4/\sqrt{\pi})(V_0/E_R)^{3/4}\exp[-2\sqrt{V_{0}/E_{\text{R}}}]$ \cite{bloch08}.
Realistic parameter values [see below for details] suggest that the
low-temperature, strong-interaction regime
$U\approx10t_c \gg t_c \gg k_B T \approx 1\mu$eV
lies within reach with state-of-the-art experimental techniques.
%% %%%%%%%%%%%%%%%%%%%%%%%% FIGURE 2 %%%%%%%%%%%%%%%%%%%%%%%%%%%%%%%%%%%
\begin{figure}[t!]
\includegraphics[width=1.05\columnwidth]{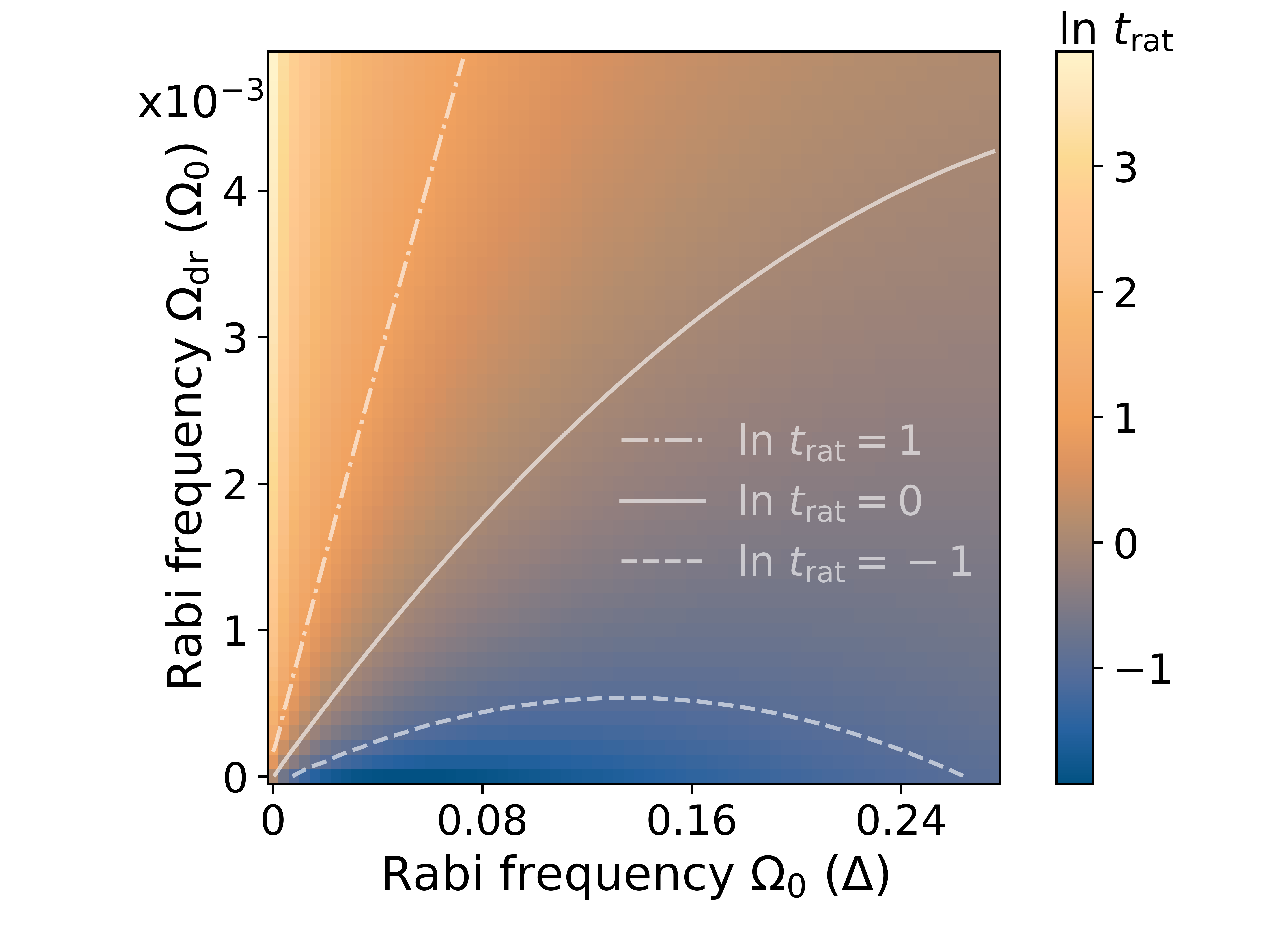}
\caption{\label{fig:tunneling}(color online).
Overview of $\log t_\mathrm{rat}$ as a function of $\Omega_\mathrm{dr}/\Omega_0$
and $\Omega_0/\Delta$.
The contour lines depict parameter constellations of equal $t_\mathrm{rat}$:
$t_\mathrm{rat} = 10$ (dash-dotted), $t_\mathrm{rat} = 1$ (solid),
$t_\mathrm{rat} = 0.1$ (dashed).
Other parameters: $n_b=1$.
}
\end{figure}
%% %%%%%%%%%%%%%%%%%%%%%%%% FIGURE 2 %%%%%%%%%%%%%%%%%%%%%%%%%%%%%%%%%%%

As illustrated in Fig.~\ref{fig:setup}(b), the standing-wave field
distribution, as described by Eq.~(\ref{eq:model}), gives rise to
spatially separated traps for the different spin components.
Hence, adjacent potential minima host two different spin states
$\ket{+}_{\theta(\hat z)}$ and $\ket{-}_{\theta(\hat z)}$, respectively.
As a consequence, spin-flip assisted tunneling $\sim t_\pm$ between neighbouring
lattice sites is strongly suppressed, whereas next-nearest neighbours,
occupying the same internal state, are coupled much more strongly via
direct tunneling $\sim t_c$, as captured by Eq.~\eqref{eq:hubbard}.
In order to control these hopping matrix elements independently, we
consider the application of an additional magnetic driving field at
frequency $\omega_2 \neq \omega$ which effectively couples different
spin states (at adjacent lattice sites), thus increasing the hopping parameter
$t_\pm$ and at the same time also the ratio $t_\mathrm{rat} := t_\pm/t_c$.
As outlined in Appendix \ref{app:hubbard}, this introduces a second hopping
term to the Fermi-Hubbard
model in Eq.~\eqref{eq:hubbard} and the resulting Hamiltonian can be written
in a suitable co-rotating frame as
%~ %% %%%%%%%%%%%%%%%%%%%%%%%% EQUATION 5 %%%%%%%%%%%%%%%%%%%%%%%%%%%%%%%%%%%
\begin{eqnarray}\label{eq:hubbard2}
H_{\text{FH2}} & = & -t_c\sum_{\langle\langle i,j\rangle\rangle,s}(c_{is}^{\dagger}c_{js}+\text{h.c.})
-t_\pm\sum_{\langle i,j\rangle,s}(c_{is}^{\dagger}c_{j\bar s}+\text{h.c.}) \nonumber \\
 &  & + \sum_{i} \mu_i n_i + \sum_{s,s^\prime} \sum_{ijkl} U_{ijkl} c_{is^{\prime}}^{\dagger}c_{js}^{\dagger}c_{ls}c_{ks^{\prime}},
\end{eqnarray}
%~ %% %%%%%%%%%%%%%%%%%%%%%%%% EQUATION 5 %%%%%%%%%%%%%%%%%%%%%%%%%%%%%%%%%%%
where $s$ and $\bar s$ denote opposite spins (i.e., $s =+$, $\bar s =-$ or vice versa).

The additional transverse driving field of strength $\sim \Omega_\mathrm{dr}$
has to be sufficiently small in order to be considered a perturbation
to the magnetic-lattice Hamiltonian in Eq.~\eqref{eq:adiabatic-model};
more precisely, we demand $\Omega_\mathrm{dr} \ll \Omega_0$.
In general, the time-dependence and exact form of this spatially homogeneous
field can be derived and reverse-engineered from the desired Hamiltonian
in the adiabatic frame, see Appendix \ref{app:hubbard} for further details.
Since, in the tight-binding regime, next-nearest neighbour hopping is
exponentially suppressed, weak driving fields $\Omega_\mathrm{dr}/\Omega_0 \ll 1$
are sufficient to reach situations where $t_\pm \gtrsim t_c$ and, typically,
for moderate driving strengths direct tunneling processes $\sim t_c$ can be safely neglected \cite{jaksch98}.
In Fig.~\ref{fig:tunneling}, it is shown how the ratio $t_\mathrm{rat}$
is affected by sweeping $\Omega_\mathrm{dr}/\Omega_0$ and
$\Omega_0/\Delta$, while keeping the number of bound states $n_b\approx\sqrt{V_0/(4E_R)}$
at a constant value.
Evidently, smaller driving fields $\Omega_\mathrm{dr}$ lead to smaller
$t_\pm$.
Moreover, at small $\Omega_0/\Delta \ll 1$ (i.e.~deep in the perturbative
regime, see Sec.~\ref{ssec:single-particle}), $t_\mathrm{rat}$ tends to
decrease with increasing $\Omega_0/\Delta$.
By choosing adequate driving fields, the tunneling matrix elements $t_c$
and $t_\pm$ can thereby be independently tuned over a relatively wide range.

\textit{Spin-orbit interaction}.\textemdash
In the presence of strong spin-orbit interaction (SOI),
transitions between different spin states at adjacent
lattices sites can be induced (eventually, for strong enough SOI,
without any external driving field) such that the Hubbard model in
Eq.~\eqref{eq:hubbard} may contain additional SOI-induced hopping terms.
Specifically, SOI-induced hopping parameters can be estimated as
$t_\pm^\lambda/E_\text{R} \approx \lambda \sqrt{V_0 E_\text{R}}
\pi^2 / a \exp \left ( -\pi^2/16 \sqrt{V_0 / E_\text{R}} \right )$,
where $\lambda = \alpha_\mathrm{R}, \beta_\mathrm{D}$ denotes the Rashba and Dresselhaus coupling
strengths, respectively.
For realistic parameter values, this may give rise to $t_\pm^\lambda / t_c \gtrsim 1$
such that nearest and next-nearest neighbour hopping terms become comparable,
see Sec.~\ref{sec:case-study} for further details.
Both the Rashba and Dresselhaus SOI strengths depend on the orientation
of the lattice in the host material and can thereby induce anisotropic
hopping.
This gives access to a wider class of Hubbard models than those captured by
Eq.~(\ref{eq:hubbard2}).

%% %%%%%%%%%%%%%%%%%%%%%%%%%%%%%%%%%%%%%%%%%%%%%%%%%%%%%%%%%%%%%%%%%%%%%

%% %%%%%%%%%%%%%%%%%%%%%%%%%%%%%%%%%%%%%%%%%%%%%%%%%%%%%%%%%%%%%%%%%%%%%
%% Section 3:
%% IMPLEMENTATIONS
\section{Implementations \label{sec:implementations}}

In the following, we propose two experimental setups for the realization
of our model.
First, in Sec.~\ref{ssec:LER}, we consider magnetic field gradients provided
by a classical current source as an example for
a setup which can be operated both in a static ($\omega = 0$; compare
Eq.~(\ref{eq:model})) or dynamic ($\omega > 0$) mode.
Subsequently, we will discuss a purely dynamic (i.e., always $\omega > 0$)
setup based on surface acoustic waves in Sec.~\ref{ssec:SAWs}.

%%% Implementation I: LER
\subsection{Superconducting circuit \label{ssec:LER}}
\begin{figure}[b!]
\includegraphics[width=1\columnwidth]{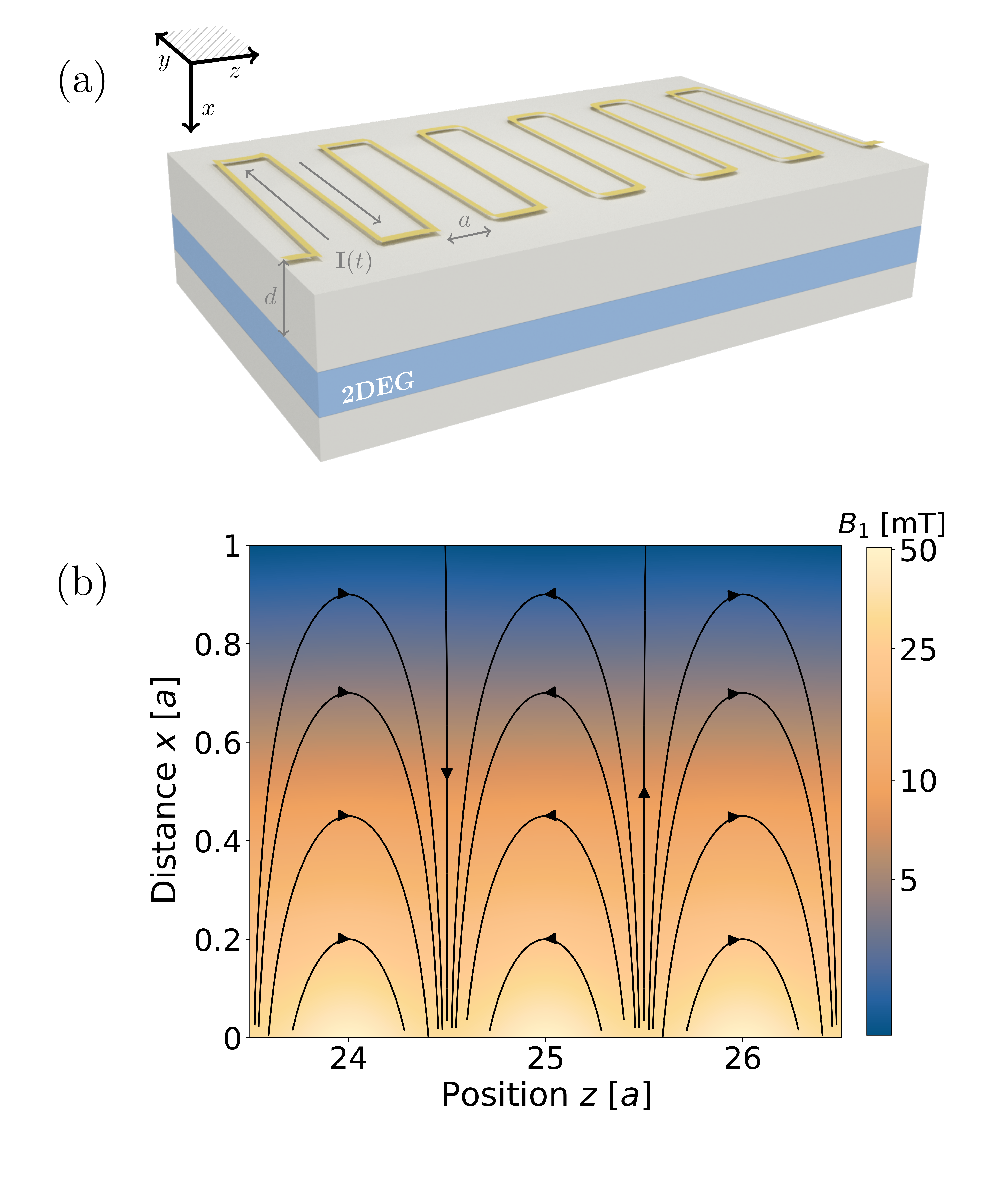}
\caption{\label{fig:LER-setup}(color online).
(a) Sketch of the meandering-wire setup.
A current provides a magnetic field as described by the Biot-Savart law.
At a distance $x = d$ from the surface, the two-dimensional electron gas
is located (see text).
(b) Magnetic field distribution for an example of a meandering nanowire that consists
of $N = 50$ parallel wires which are separated by the lattice constant $a = 1~\mu$m.
The vector field $\mathbf{B}_\mathrm{AC}(\mathbf{r},t=0)$ is shown and
its scalar field
$|\mathbf{B}_\mathrm{AC}|$ is plotted on a
logarithmic scale.
Magnetic field strenghts of the order of $B_1 \sim (10-50)~$mT are
obtained in the proximity ($x \lesssim 0.6 a=600~$nm) of the wire.
Other numerical parameters: $I_0 = 70 ~ \mathrm{mA}$ at a current density
$J_\mathrm{c} = 30~$MA/cm${}^2$ \cite{ilin14} and wire dimensions of $480~$nm x $480~$nm.
}
\end{figure}

As a first example for a realization of our model as described by Eq.~(\ref{eq:model}),
we consider SC circuits operating at GHz frequencies.
The electrons are confined in a 2DEG at a distance $d$ from a current-carrying wire,
which is located above the surface.
For our purposes, SC circuits and circuit resonators are
attractive because of
their capability to generate AC magnetic fields by carrying relatively
large currents and the possibility to integrate them in semiconductor
nanostructures \cite{tosi14,sarabi17}.
In a simple toy model, we describe the circuit by a meandering wire carrying
an AC current $\sim I_0 \cos(\omega t)$ through parallel sections of the
wire separated by a lattice constant $a$, see Fig.~\ref{fig:LER-setup}(a)
for an illustration of the setup.
Note that, in principle, this setup can also be operated in the static
regime ($\omega = 0$) when DC currents and, thus, time-\textit{independent}
fields are considered.
The classical electric current density $\mathbf{J}$ induces
a magnetic field which is calculated using the Biot-Savart law,
see Fig.~\ref{fig:LER-setup}(b) for an exemplary field distribution
as induced by a current source at fixed positions $\mathbf{r} = (0<x<a,y=0,23.5<z/a<26.5)$
\cite{footnote1}.

%% %%%%%%%%%%%%%%%%%%%%%%%% FIGURE 3 %%%%%%%%%%%%%%%%%%%%%%%%%%%%%%%%%%%
%% %%%%%%%%%%%%%%%%%%%%%%%% FIGURE 3 %%%%%%%%%%%%%%%%%%%%%%%%%%%%%%%%%%%

Here, we consider only one-dimensional trapping potentials in which the
electrons are confined to a one-dimensional channel such that the $y$
motional DOF is frozen out.
Furthermore, we assume that the spatial extension of the meandering wire
exceeds the size of the trapping region within the 2DEG, such that
finite-size effects of the induced magnetic field can be neglected.
This simplifies the mathematical description and we obtain the
AC magnetic field distribution $\mathbf{B}_\text{AC}(\mathbf{r}, t)=\Omega(\mathbf{r})\cos(\omega t) \hat{\mathbf{x}}$
for a given wire geometry by summing up the induced fields of all
parallel wire segments,
see Fig.~\ref{fig:LER-setup} [for details, cf.~Appendix \ref{app:ler}].
In the presence of an additional static homogeneous field
$\mathbf{B}_\text{ext} = B_\text{ext} \hat{\mathbf{z}}$,
the resulting Hamiltonian, $H(t) = \hat p^2/(2m) + \gamma (\mathbf{B}_\text{AC}(\mathbf{\hat r},t) + \mathbf{B}_\text{ext}) \cdot \bm{\sigma}$,
approximately coincides with our model in Eq.~(\ref{eq:model}),
where we can identify $\omega_0 = \gamma B_\text{ext}$
and the amplitude $\Omega_0$ of the Rabi frequency is given by
%% %%%%%%%%%%%%%%%%%%%%%%%% EQUATION 6 %%%%%%%%%%%%%%%%%%%%%%%%%%%%%%%%%
\begin{equation}\label{eq:Rabi-LER}
\Omega_0 = \gamma \frac{\mu_0 I_0 d}{\pi a^2} \sum_{n \in \mathbb{N}_0} \frac{(-1)^n}{(n+\frac{1}{2})^2 + \left ( \frac {d}{a} \right )^2}.
\end{equation}
%% %%%%%%%%%%%%%%%%%%%%%%%% EQUATION 6 %%%%%%%%%%%%%%%%%%%%%%%%%%%%%%%%%
Eq.~(\ref{eq:Rabi-LER}) becomes exact in the limit of an infinitely
long wire and it converges to the numerically exact result in the limit
of a long wire and in the center region below the wire
[see App.~\ref{app:ler}
for further details]; for all practical purposes, it yields sufficiently
exact results for typical resonator geometries.
The exact spatial pattern of the Rabi frequency $\Omega(\hat z)$ depends on both
the geometry of the resonator and the ratio $d/a$.
Neglecting finite-size effects and for a perfectly periodic resonator
geometry, the Rabi frequency can be approximately written as
$\Omega(\hat z) = \Omega_0 \cos(\pi \hat z / a +\phi)$, see Appendix \ref{app:ler} for
further details.

Let us conclude the description of the proposed setup
with a few general remarks.
Firstly, we note that the calculation of the Hamiltonian results in
an additional time-dependent term $\propto \sigma^z$ which we have neglected
here and which is typically very small compared to the time-independent
contribution from $B_\text{ext}$, see Appendix \ref{app:ler} for
more information.
Secondly, the calculated RF field strength $B_1 \approx (10-50)$~mT [see Fig.~\ref{fig:LER-setup}(b)]
at a given distance $d \lesssim 0.6a$
and given current intensity $I_0 = 70~\mathrm{mA}$ from the surface ranges
from realistic to very optimistic values.
The highest given values can only be obtained in close proximity to the
surface.
Moreover, the critical current density $J_\mathrm{c} = 30~$MA/cm${}^2$ \cite{ilin14}
used in our calculations is optimistic because high ($\sim$ GHz) frequencies
and strong ($\sim$ T) in-plane magnetic fields might reduce this value.
However, especially the frequency dependence of $J_\mathrm{c}$ is still a current
topic of research and, as noted earlier, the proposed setup may also be
operated at $\omega=0$, i.e., with DC currents.
For $g$-factors $\sim 15$ (e.g., in InAs-based quantum wells), the given
range of field strengths amounts to trap depths
$V_0 \lesssim (4-22)~\mu\mathrm{eV} = k_\text{B} \cdot (46-255)~$mK.
An explicit case study for specific material parameters follows in
Sec.~\ref{sec:case-study}, where we check when the requirements
set by Eq.~(\ref{eq:requirements}) can be fulfilled.
Finally, we stress that the relevant system parameters from
Eq.~(\ref{eq:requirements}) do not depend on the material choice
(except for the $g$-factor of the quantum well) and due to its simplicity,
the setup can, in principle, readily be implemented in an experiment.
%%%
While the trap depth $V_0$ is tunable, the geometry is predefined in
this setup, and therefore the lattice constant $a$ (thus also the ratio
$d/a$) is fixed.
In the following, we will discuss an implementation which overcomes this
limitation by construction, allowing for more widely tunable system
parameters and lattice geometries.
%%%

%%% Implementation II: SAWs
\subsection{Surface acoustic waves \label{ssec:SAWs}}

As a second implementation, we discuss time-dependent ($\omega > 0$)
magnetic field gradients induced by SAWs.
In piezomagnetic materials which exhibit a significant (inverse)
magnetostrictive effect, mechanical and magnetic DOFs are coupled which
can be captured by the constitutive relations for magnetostriction,
cf.~Appendix \ref{app:SAWs}.
Specifically, the magnization $\mathbf{m}$ of a sample with non-zero
magnetoelastic coupling changes due to mechanical stress applied to the
material, which is described by a stress tensor $\underline{\underline{T}}$.

\begin{figure}[t!]
  \centering
\includegraphics[width=1\columnwidth]{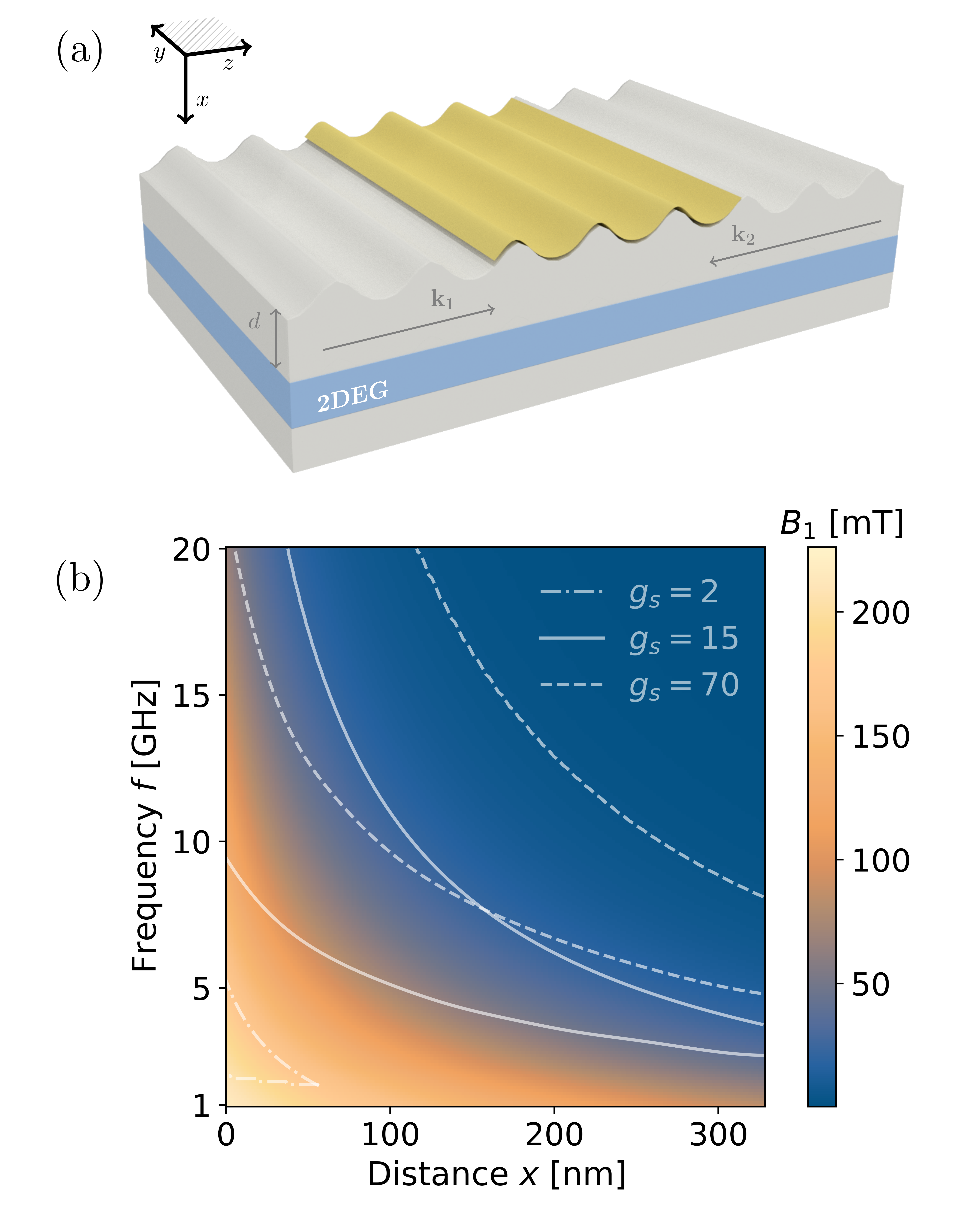}
\caption{
(a) Sketch of the SAW-based setup with a ferromagnetic film above the surface.
Two counter-propagating SAWs generate standing-wave mechanical and
magnetic field distributions.
(b) Magnetic field strength $B_1$ as a function of distance $x$ from
the ferromagnetic film and SAW frequency $f$.
The contour lines indicate the regions where
$k_\mathrm{B} T = 1\mu\mathrm{eV} \ll \gamma B_1/2 \ll 2\pi f$
[see Eq.~\eqref{eq:requirements}]
can be fulfilled for different $g$-factors: $g_\text{s} = 2$ (dash-dotted lines),
$g_\text{s} = 15$ (solid lines), $g_\text{s} = 70$ (dashed lines).
Other numerical parameters: Speed of sound $v_\text{s} = 3500$~m/s,
film thickness $\delta = 25~$nm, saturation magnetization
$\mu_0|\mathbf{m}_\text{s}| = 1.8~T$, strain amplitude
$\varepsilon_{xx} = 2 \cdot 10^{-4}$, damping constant $\alpha = 0.01$,
magnetoelastic constant $h = 10~T$, g-factor of the ferromagnetic film $g_\text{s,FM} = 2.1$.
}
\label{fig:setup-SAWs}
\end{figure}

We consider a ferromagnetic film of thickness $\delta$ deposited on top of a SAW-carrying substrate,
where the surface waves generate RF strain fields which, in turn, can induce
magnetization dynamics in the ferromagnet and may thus provide strong time-dependent
magnetic stray fields;
for related experimental works, see Refs.~\cite{dreher12,salasyuk17}.
This setup is schematically shown in Fig.~\ref{fig:setup-SAWs}(a).
Two counter-propagating SAWs, which can be launched from interdigital
transducers (IDTs) patterned on top of the material, generate a standing-wave pattern of both
the mechanical field and induced spin wave, introducing a periodicity
which defines the lattice constant $a=\lambda/2$ where $\lambda$ is the
SAW wavelength;
the dispersion relation of the SAW, $\omega=2\pi f=kv_s$, yields $\lambda=v_s/f$,
where $v_\text{s}$ denotes the speed of sound in the host material.
This results in a spatially and time-periodic magnetic field as needed for the realization
of Eq.~(\ref{eq:model}).
The coupled equations of motion for the (\textit{i}) mechanical
and (\textit{ii}) magnetic field distributions can be described by (\textit{i})
$\rho \ddot u_i = \partial T_{ij} / \partial z_j$, where $\rho$ and
$\mathbf{u}(\mathbf{x},t)$ denote the mass density and the mechanical
displacement vector, respectively, with the displacement $u_i$ along the
coordinate $\hat z_i \ (= \hat x, \hat y, \hat z)$
and (\textit{ii}) the Landau-Lifshitz-Gilbert (LLG) equation, respectively.
The latter describes the motion of the unitless magnetization direction
$\mathbf{m}$ due to an effective magnetic field $\mathbf{H}_\text{eff}$
and reads \cite{landau35,gilbert04}
%% %%%%%%%%%%%%%%%%%%%%%%%% EQUATION 7 %%%%%%%%%%%%%%%%%%%%%%%%%%%%%%%%%%%
\begin{equation}\label{eq:LLG}
\frac{\partial \mathbf{m}}{\partial t} = - \gamma \mathbf{m} \times \mu_0 \mathbf{H}_\text{eff} + \alpha \mathbf{m} \times \frac{\partial \mathbf{m}}{\partial t},
\end{equation}
%% %%%%%%%%%%%%%%%%%%%%%%%% EQUATION 7 %%%%%%%%%%%%%%%%%%%%%%%%%%%%%%%%%%%
where $\mu_0$ and $\alpha$
denote the magnetic constant and phenomenological Gilbert damping parameter,
respectively, and $\mathbf{H}_\text{eff}$ accounts for the SAW-induced
magnetic field.

%% %%%%%%%%%%%%%%%%%%%%%%%% FIGURE 4 %%%%%%%%%%%%%%%%%%%%%%%%%%%%%%%%%%%
%% %%%%%%%%%%%%%%%%%%%%%%%% FIGURE 4 %%%%%%%%%%%%%%%%%%%%%%%%%%%%%%%%%%%

Given the effective magnetic field $\mathbf{H}_\text{eff}$ at the
ferromagnetic film ($x=0$) which is calculated from Eq.~(\ref{eq:LLG}), we
estimate the stray field at the 2DEG, see Appendix \ref{app:SAWs} for details.
The accessible range of field strenghts $B_1$ strongly depends on the
specific material-dependent parameters, i.e., the saturation magnetization
$\mathbf{m}_\text{s}$, the damping parameter $\alpha$, the $g$-factor
$g_\text{s,FM}$ and magnetoelastic constant $h$ of the film and, moreover,
the amplitude of the SAW-induced strain field.
The latter is technically limited due to undesired heating effects at
too large amplitudes.
Fig.~\ref{fig:setup-SAWs}(b) shows the RF field strength $B_1$
as a function of distance $x$ from the ferromagnetic film and SAW
frequency $f$.
The numerical parameters are chosen such that they can be implemented
in state-of-the-art experiments [see caption of Fig.~\ref{fig:setup-SAWs}];
note that even much higher strain amplitudes \cite{sherman13}, magnetoelastic
constants \cite{dreher12} and lower damping constants \cite{schoen16} have been
realized in experiment, which renders our chosen set of parameters
very realistic.
As a result, we obtain strong driving fields $B_1 \approx (10-100)$~mT
at given distance $x = (0.1 - 0.5)a$ from the film which amounts to
trap depths $V_0 \lesssim (4-43)~\mu$eV at $g_\text{s} \sim 15$.
However, for increasing frequencies $f \sim (10 - 50)$~GHz, the field
strength decreases at fixed distance $x$.
Hence, the lattice constant cannot be made arbitrarily small.
In Sec.~\ref{sec:case-study}, we provide an overview of realistic
parameter regimes (specifically, with a focus on Eq.~(\ref{eq:requirements}))
based on the derived driving fields.

\textit{Strain-induced acoustic traps}.\textemdash
So far, we have neglected strain-induced deformation potentials and
electric-field components generated in a piezoelectric host material.
In principle, these electric fields couple to the motional DOF of a
charged particle and thereby induced time-dependent electric potentials
can either constitute stable traps or, if the driving amplitude of the
electric field becomes too large, destabilize the motion of the electron
\cite{schuetz17}.
In order to take both the electric and magnetic field-induced couplings
to the external \textit{and} internal DOFs into account, we extend our
previous analysis to the more general model
%% %%%%%%%%%%%%%%%%%%%%%%%% EQUATION 8 %%%%%%%%%%%%%%%%%%%%%%%%%%%%%%%%%%%
\begin{eqnarray}\label{eq:hamilt-AL-and-ML}
H_\mathrm{hyb} & = & \frac{\hat p^2}{2m} + V_\text{SAW} \cos(k \hat z) \cos (\omega t) \nonumber \\
& & + \frac{\omega_0}{2} \sigma^z + \frac{\Omega_0}{2} \cos ( k \hat z ) \cos(\omega t) \sigma^{x},
\end{eqnarray}
%% %%%%%%%%%%%%%%%%%%%%%%%% EQUATION 8 %%%%%%%%%%%%%%%%%%%%%%%%%%%%%%%%%%%
which contains a kinetic term, a time-dependent strain-induced potential
of amplitude $V_\text{SAW}$ and the remaining terms from the Hamiltonian in Eq.~(\ref{eq:model}).
Following the procedure outlined in Refs.~\cite{rahav03,rahav03b}, we derive an effective
time-independent Hamiltonian for the hybrid (strain-induced and magnetic)
lattice by performing a high-frequency expansion of Eq.~\eqref{eq:hamilt-AL-and-ML}
in $1/\omega$.
Starting from Eq.~\eqref{eq:hamilt-AL-and-ML}, we obtain an effective
model of the form
\begin{equation}\label{eq:hamilt-AL-and-ML-effective}
H_\mathrm{hyb}^\text{eff}=\frac{\hat p^2}{2m} + \frac{|\Delta|}{2} \tilde \sigma^z +
\left [ \frac{V_\text{SAW}^2}{8E_\mathrm{S}} - \frac{\Omega_0^2}{4|\Delta|} \tilde \sigma^z \right ] \sin^2(k \hat z),
\end{equation}
with $E_\text{S} = m v_s^2/2$.
This result can be self-consistently verified
in the limit $\Omega_0/|\Delta|,V_\mathrm{SAW}^2/(8E_\mathrm{S}^2)\ll1$.
The second term in Eq.~\eqref{eq:hamilt-AL-and-ML-effective} describes
a spin-dependent energy offset [compare Fig.~\ref{fig:setup}] and the third
term is a spin-dependent effective potential.

\begin{figure}[t!]
  \centering
   \includegraphics[width=0.5\textwidth]{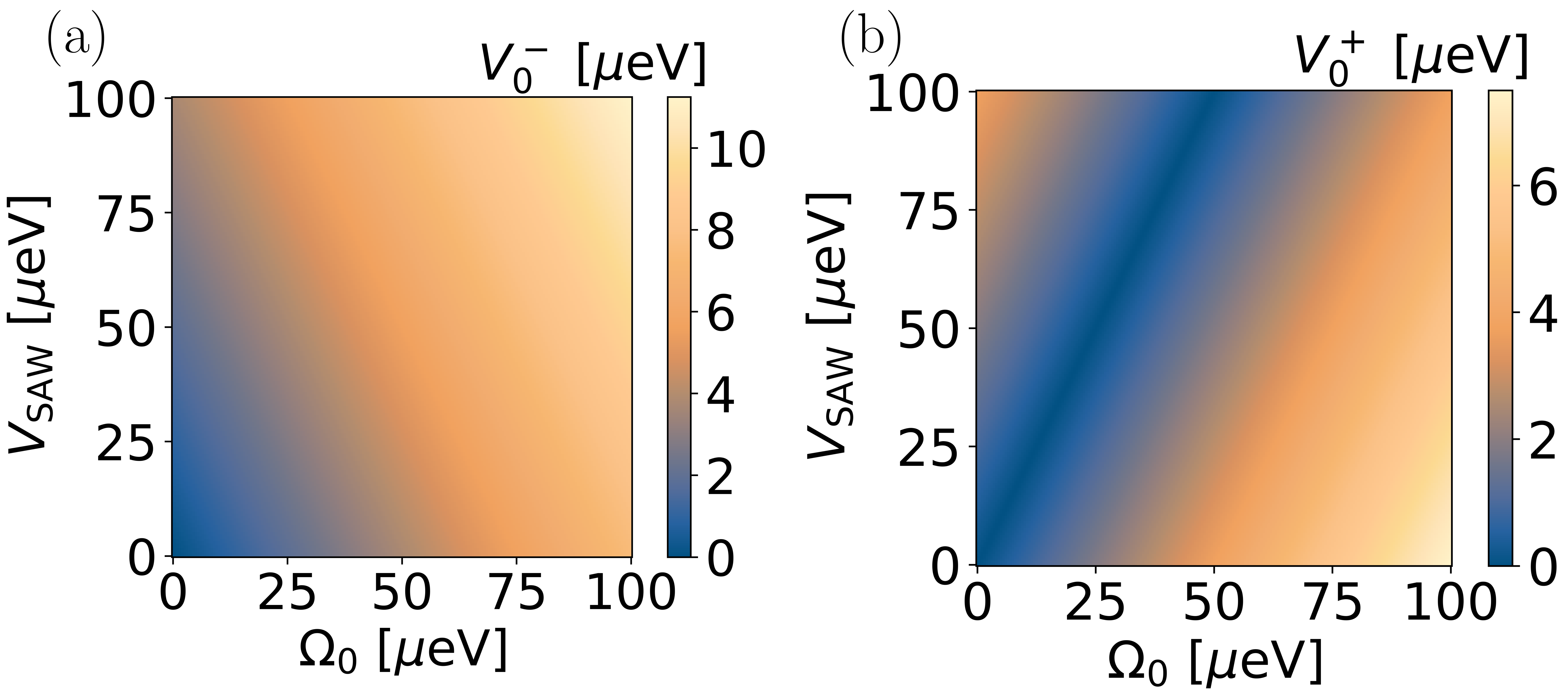}
\caption{
Spin-dependent trap depth of effective potential as given by
Eq.~\eqref{eq:hamilt-AL-and-ML-effective} plotted as a function of
Rabi frequency $\Omega_0$ and strain-induced potential amplitude $V_\mathrm{SAW}$
for fixed $\Omega_0/|\Delta|=0.3$ and $V_\mathrm{SAW}/E_\mathrm{S}=0.3$.
(a) Effective trap depth of hybrid trap for the $s=-$ spin component.
The magnetic and strain-induced potentials add up and the effective potential
becomes deeper if either the magnetic or strain contribution is increased.
(b) Effective trap depth of hybrid trap for the $s=+$ spin component.
The magnetic and strain-induced potentials have different signs.
At $V_\mathrm{SAW}=2\Omega_0$, the two potentials cancel each other.
}
\label{fig:hybrid-potential}
\end{figure}

From Eq.~\eqref{eq:hamilt-AL-and-ML-effective}, by projecting onto the
adiabatic eigenstates $\ket{+}_{\theta(\hat z)}$ and $\ket{-}_{\theta(\hat z)}$,
respectively, we obtain the spin-dependent potential amplitudes, i.e.,
$V_0^- = \Omega_0^2/(4|\Delta|) + V_\text{SAW}^2/(8E_\mathrm{S})$
and
$V_0^+ = |\Omega_0^2/(4|\Delta|) - V_\text{SAW}^2/(8E_\mathrm{S})|$.
We can deduce that strain-induced and magnetic
potentials add up constructively (destructively) for the $\ket{-}_{\theta(\hat z)}$
($\ket{+}_{\theta(\hat z)}$) adiabatic potential.
In Fig.~\ref{fig:hybrid-potential} the effective trap depths for both
spin components are shown as a function of $\Omega_0$ and $V_\mathrm{SAW}$.
Since the strain-induced deformation potential is typically very weak \cite{hanson07,naber06,schuetz15},
we consider the strain-induced potential $\sim V_\mathrm{SAW}$ to become
important only in piezoelectric materials.
However, since the magnetic traps operate at relatively high strain amplitudes
[see Sec.~\ref{ssec:SAWs}], in piezoelectric materials this contribution
will typically not be negligible and also depends on the orientation of
the magnetic lattice with respect to the crystalline structure of the
piezoelectric host medium.
More details on the derivation of Eq.~\eqref{eq:hamilt-AL-and-ML-effective}
and a stability analysis of the time-dependent model Hamiltonian given in
Eq.~\eqref{eq:hamilt-AL-and-ML} can be found in Appendix \ref{app:SAWs}.
%% %%%%%%%%%%%%%%%%%%%%%%%% FIGURE 5 %%%%%%%%%%%%%%%%%%%%%%%%%%%%%%%%%%%
%% %%%%%%%%%%%%%%%%%%%%%%%% FIGURE 5 %%%%%%%%%%%%%%%%%%%%%%%%%%%%%%%%%%%

%% %%%%%%%%%%%%%%%%%%%%%%%%%%%%%%%%%%%%%%%%%%%%%%%%%%%%%%%%%%%%%%%%%%%%%
%% Section 4:
%% Case study
\section{Case studies \label{sec:case-study}}

%% %%%%%%%%%%%%%%%%%%%%%%%% TABLE 1 %%%%%%%%%%%%%%%%%%%%%%%%%%%%%%%%%%%%
\begin{table}
\begin{tabular}{l||c|c|c|}
host material & $|g_\mathrm{s}|$ & $\Omega_0^\mathrm{wire} [\mu\text{eV}]$ & $\Omega_0^\mathrm{SAW} [\mu\text{eV}]$\tabularnewline
\hline 
\hline 
GaAs & 0.44 & $\sim(0.3-1.3)$ & $\sim(1.3-2.5)$\tabularnewline
\hline 
InAs & 14.9 & $\sim(8.6-43)$ & $\sim(43-86)$\tabularnewline
\hline 
InSb & $\sim 70$ & $\sim(41-200)$ & $\sim(200-410)$\tabularnewline
\hline 
DMS & $\sim(10^2-10^3)$ & $\sim(58-2900)$ & $\sim(290-5800)$\tabularnewline
\hline 
$\text{MoS}_2$ & 2.21 & $\sim(1.3-6.4)$ & $\sim(6.4-13)$\tabularnewline 
\hline 
$\text{WS}_2$ & 2.84 & $\sim(1.6-8.2)$ & $\sim(8.2-16)$\tabularnewline 
\hline 
\end{tabular}
\caption{\label{tab:case-study}
Estimates for achievable Rabi frequencies in both the nanowire and SAW setups.
The table shows Rabi frequencies based on both state-of-the-art
($B_1^\text{wire} = 10$mT, $B_1^\text{SAW} = 50$mT)
and more optimistic
($B_1^\text{wire} = 50$mT, $B_1^\text{SAW} = 100$mT)
maximum driving field strengths [compare Figs.~\ref{fig:LER-setup} and \ref{fig:setup-SAWs}].
}
\end{table}
%% %%%%%%%%%%%%%%%%%%%%%%%% TABLE 1 %%%%%%%%%%%%%%%%%%%%%%%%%%%%%%%%%%%%
\textit{Faithful implementation of magnetic traps}.\textemdash
As outlined above, a faithful implementation of magnetic traps is only
possible if Eq.~(\ref{eq:requirements}) can be fulfilled.
This can be achieved in state-of-the-art experiments, e.g., in the setups
discussed in Sec.~\ref{sec:implementations}, as we outline in the following:
(\textit{i}) The spontaneous phonon emission rate can be as low as $\gamma \sim 0.3~\mu$eV
in InAs-based setups \cite{liu14} and similar values are expected for InSb-based
setups \cite{sousa03}.
Even for much higher emission rates, the regime $\gamma \ll \omega_\text{HO}$
can still be reached and, typically, $k_\text{B} T \approx (1-10)~\mu\mathrm{eV}\ll\omega_\mathrm{HO}$
imposes a stronger constraint on the minmum energy $\omega_\text{HO}$.
(\textit{ii}) Based on the results shown in Figs.~\ref{fig:LER-setup} and \ref{fig:setup-SAWs},
Table \ref{tab:case-study} gives an overview of realistic Rabi frequencies $\Omega_0$
in both described setups for different host materials \cite{footnote3}.
Since the trap depth $V_0 \lesssim \Omega_0 /2$ is limited from above by
half of the Rabi frequency $\Omega_0$, it is evident that relatively
low-$g_\text{s}$ materials, like, e.g., GaAs, do not prove to be promising
candidates for magnetic trapping as described in Sec.~\ref{sec:theory}
since, in particular, the condition $k_\text{B}T\ll V_0 \lesssim \Omega_0/2$
from Eq.~\eqref{eq:requirements} cannot be fulfilled easily.
Assuming thermal energies $k_\mathrm{B} T \approx (1-10)~\mu$eV, a comparison
with the data shown in Table \ref{tab:case-study} suggests that a faithful
implementation of magnetic traps should be feasible with state-of-the-art
experiments using materials with moderate (e.g., TMDs like $\text{MoS}_2$
or $\text{WS}_2$) to relatively high g-factors $|g_\text{s}| \gtrsim 15$
(as can be found, e.g., in III-V semiconductors like InAs or InSb).
Only then, thermal stability as required by Eq.~\eqref{eq:requirements}
can be guaranteed.
(\textit{iii}) Given that trap depths of the order of $V_0 \sim 100~\mu$eV may be
reached in SAW-based setups at $|g_\text{s}| \gtrsim 15$, the requirements
$k_\text{B} T \ll \omega_\text{HO} \lesssim V_0<\omega/2$ can be fulfilled at
oscillator frequencies $\omega_\text{HO} \gtrsim 5~\mu$eV ($\gtrsim 7.5~$GHz).
In this parameter regime, accordingly, the trap can support a couple of
bound states $n_b \approx 1-5$.
(\textit{iv}) Moreover, as discussed in detail in Sec.~\ref{ssec:single-particle},
high driving frequencies $f=\omega/(2\pi) \gtrsim 10$~GHz are another important
bottleneck towards the experimental realization of reliable magnetic traps;
these can be provided by both the proposed nanowire and SAW-based setups,
as has been experimentally demonstrated, reaching ultra-high frequencies
$f \approx 25~$GHz ($\omega \approx 103~\mu$eV) \cite{kukushkin04}.
Using existing technology, as indicated, e.g., by the solid lines in
Fig.~\ref{fig:setup-SAWs}, experiments could therefore be operated in a
regime where $\Omega_0\lesssim\omega$ (and even the more demanding
requirement (within RWA) $\Omega_0\ll\omega$) is clearly fulfilled.
(\textit{v}) 2DEGs in InAs-based quantum wells can have a long mean-free
path of the order of a few $\mu$m \cite{koester96,yang02} which is
much larger than a lattice spacing of a few hundred nm.
This provides optimism that disorder may not become too large in some of
the high $g$-factor materials considered here, cf.~also Ref.~\cite{schuetz17}
for a more detailed discussion on the role of disorder in related systems.

\textit{Parameter regimes for Fermi-Hubbard physics in magnetic lattices}.\textemdash
Typical tunneling rates $t_c$ in magnetic lattices (as described in Sec.~\ref{ssec:many-body})
can reach values of a couple of $\mu$eV as discussed below.
By sufficiently screening the Coulomb interaction, e.g., with the aid of a
metallic screening layer \cite{byrnes07}, we may enter a parameter regime
where both $t_c \gg k_\text{B} T$ and $U \approx 10t_c$ can be
reached simultaneously which itself is interesting for studying phenomena
of quantum magnetism \cite{bloch08}.
Furthermore, we introduced in Sec.~\ref{ssec:many-body} the possibility
to enrich the standard Fermi-Hubbard model, typically including only tunneling
processes between adjacent lattice sites, by the application of additional
driving fields [see also Appendix \ref{app:hubbard}], thus allowing for
independent tuneability of the hopping parameters $t_c$ and $t_\pm$.
Weak driving fields $\Omega_\text{dr} \ll \Omega_0$ already give access
to all the different regimes $t_\pm\ll t_c$, $t_\pm \approx t_c$ and $t_\pm \gg t_c$.

For SOI-induced hopping process $\sim t_\pm^\lambda$, we estimate that
$t_\pm^\lambda \sim 50~\mu$eV can be reached at lattice spacings of a few
$100~$nm in InAlAs/InGaAs quantum wells
where the Dresselhaus SOI is mostly negligible \cite{koga11} and the
Rashba parameter is given by $\alpha_\mathrm{R} \approx 10^4$~m/s \cite{manchon15}.
Note that this value depends very strongly on the host material and,
naturally, in some materials both the Rashba and Dresselhaus couplings
become important which can induce significant anisotropies \cite{hanson07}.
Most notably, this shows that the parameter regime $t_\pm^\lambda \gtrsim t_c$
is accessible and the next-nearest neighbour tunneling processes may become
important even without the application of any additional driving fields.

Within our tight-binding model where we consider the limit $V_0\gg E_\text{R}$,
$\omega_\text{HO}$ is typically of the order of a few recoil energies \cite{bloch08}.
Considering, e.g., InAs or InSb as host materials, the effective electron
mass becomes relatively small, i.e., $m_\text{InAs}=0.023m_0$ and $m_\text{InSb}=0.014m_0$,
both expressed in terms of the electron's rest mass $m_0$ \cite{singleton01}.
Then, only relatively large lattice spacings $a \gtrsim 1~\mu$m give
rise to small recoil energies $E_\text{R}\ll V_0$.
In turn, much smaller lattice spacings $a \gtrsim 300$~nm can be self-consistently
achieved in TMD-based setups, where, e.g., $m_\text{MoSe2}=0.67m_0$.

\textit{Spin relaxation and dephasing}.\textemdash
The specific value for the spin relaxation time $T_1$ is material-dependent. 
Generically, however, $T_1$ can be very long ($T_1 \sim 10~\mathrm{ms}$),
as is well known from spin relaxation measurements in quantum dots \cite{elzerman04, amasha08}. 
Therefore, on the relevant timescales considered here, spin relaxation can
be largely neglected, allowing for the faithful realization of spinful
(two-species) magnetic lattices.
Only in the presence of very strong magnetic fields, care must be taken
to avoid too fast spin relaxation, since $1/T_1 \sim B_0^5$ \cite{khaetskii01}.
Conversely, spin dephasing times $\sim T_{2}^{\star}$ tend to be much
shorter than $T_{1}$.
In InAs \cite{nadj-perge10} and InSb \cite{berg13}, e.g., values of
$T_{2}^{\star}\sim 10$~ns have been reported. 
While spin dephasing should not affect our ability to magnetically \textit{trap}
single electrons, the observation of coherent (many-body) spin physics
may be severely limited by electron spin decoherence, since the many-body
wavefunction of $N$ electrons will dephase on a timescale set by $\sim T_{2}^{\star}/N$.

\textit{Specific examples: InAs and InSb}.\textemdash
Finally, we discuss the full set of relevant system parameters for two
specific material choices, i.e., InAs-based and InSb-based setups.
In the following, we assume dilution-fridge temperatures $T=10~$mK,
i.e., $k_\text{B}T\approx1~\mu$eV.
Hence, the spontaneous phonon emission rate given above fulfills $\gamma \sim 0.3~\mu\mathrm{eV} < k_\text{B}T$,
underlining that a low $\gamma$ is expected to set the smallest energy scale in
Eq.~\eqref{eq:requirements} if thermal stability ($k_\text{B} T \ll \omega_\text{HO}, V_0$) is ensured.
First, we consider electrons in InAs with an effective mass $m=0.023m_0$.
For $\Omega_0 = 86~\mu$eV [compare Table \ref{tab:case-study}] and small
detunings $|\Delta|\ll\Omega_0$, we can reach trap depths
$V_0\approx 43~\mu$eV which ensures thermal robustness of the trap at
considered temperatures.
Operating at a high frequency $f=22~$GHz, the highest energy
scale in Eq.~\eqref{eq:requirements} is set by $\omega\approx92~\mu$eV
at a lattice spacing $a=900~$nm.
For self-consistency, we check that the recoil energy is given by
$E_\text{R}\approx20~\mu$eV which means that we are not deep in the
tight-binding limit ($E_\text{R}\ll V_0)$.
Still, the tunneling parameter can be estimated as $t_c\approx 5.2~\mu$eV \cite{bloch08}.
Note that, in this setting ($|\Delta|\ll\Omega_0$), the harmonic
approximation around a local potential minimum is typically not well justified.
Secondly, we consider heavy holes in InAs with an effective mass $m=0.836m_0$.
For an ambitious Rabi frequency $\Omega_0 = 100~\mu$eV and a large detuning
$\Delta=380~\mathrm{GHz}=250~\mu$eV, we obtain a trap depth
$V_0=|\underset{z}{\mathrm{max}}\ \varepsilon(\hat z)-\underset{z}{\mathrm{min}}\ \varepsilon(\hat z)|
\approx \Omega_0/10 = 10~\mu$eV.
Operating at a high SAW frequency $f=25~$GHz, we obtain
$\omega\approx103~\mu$eV at a lattice spacing $a=500~$nm and
$v_s=25~$km/s.
Hence, the recoil energy is given by $E_\text{R}\approx1.8~\mu$eV which
ensures the validity of the tight-binding approximation.
Since the harmonic approximation, $\varepsilon(\hat z) \propto \Omega^2(\hat z) \propto \hat z^2$,
is well justified in this case, we estimate $m\omega_\text{HO}^2\hat z^2/2 \approx \Omega(\hat z)^2/(4|\Delta|)$, i.e.,
\begin{equation*}
\omega_\text{HO}=118~\mathrm{MHz} \times \sqrt{\frac{\left ( g_s [g_0]\right )^2}{m [m_0]}} \times \frac{B_1 \left [ \mathrm{mT} \right ]}{a \left [ \mu\mathrm{m} \right ] \sqrt{|\Delta [\mathrm{GHz}]|}},
\end{equation*}
where $g_0=2$ denotes the $g$-factor of the free electron.
Accordingly, we obtain $\omega_\text{HO}=5.4~\mu$eV for heavy holes in
InAs, as considered here.
Hence, all conditions imposed by Eq.~\eqref{eq:requirements} are fulfilled.
In this scenario, the tunneling parameter amounts to only $t_c\approx 0.2~\mu$eV.
However, the second hopping parameter introduced in Sec.~\ref{ssec:many-body},
$t_\pm$, can be significantly enhanced such that $t_\pm\gg t_c$ with the
aid of additional driving fields, as discussed in more detail in Appendix
\ref{app:hubbard}.
Thirdly, we consider heavy holes in InSb with an effective mass $m=0.627m_0$.
For a Rabi frequency $\Omega_0=200~\mu$eV [compare Table \ref{tab:case-study}]
and a relatively small detuning
$\Delta=38~\mathrm{GHz}=25~\mu$eV, we obtain a trap depth $V_0\approx90~\mu$eV.
Assuming a very high (SAW) frequency $f=50~$GHz, we obtain $\omega\approx207~\mu$eV
at $a=100$ nm and (in the SAW implementation) $v_s=10~$km/s.
The recoil energy is then given by $E_\text{R}\approx60~\mu$eV.
The tunneling parameter can be estimated as $t_c\approx 18~\mu$eV.

Altogether, these considerations clearly suggest that thermally stable
and well-performing magnetic traps may be implemented with current
technology; more specifically, fulfilling Eq.~\eqref{eq:requirements}
should be possible in host materials possessing high enough $g$-factors.
Furthermore, note that the values presented in Table \ref{tab:case-study}
might be further enhanced; in the SAW setup, the values calculated in
Sec.~\ref{ssec:SAWs} have been derived assuming a magnetoelastic constant
$h = 10$~T and strain amplitudes $\varepsilon_{xx}=2\cdot 10^{-4}$,
which both may be elevated further in experiment, yielding even higher
Rabi frequencies than the ones given in Table \ref{tab:case-study}.

%% %%%%%%%%%%%%%%%%%%%%%%%%%%%%%%%%%%%%%%%%%%%%%%%%%%%%%%%%%%%%%%%%%%%%%

%% %%%%%%%%%%%%%%%%%%%%%%%%%%%%%%%%%%%%%%%%%%%%%%%%%%%%%%%%%%%%%%%%%%%%%
%% Section 5:
%% SUMMARY & OUTLOOK
\section{Summary \& Outlook \label{sec:outlook}}

To summarize, we have proposed  magnetic traps and scalable lattices for
electrons in semiconductors.
Firstly, we have derived a general theoretical framework fit to characterize
the traps and parameter regimes in which they can be operated under
realistic experimental conditions and at dilution-fridge temperatures.
Secondly, we have described two possible platforms suitable for an
experimental demonstration of thermally stable magnetic traps and, eventually,
coherent lattice physics in scalable arrays of magnetic traps.
The developed model which is based on a periodically modulated AC Stark shift induced by magnetic RF fields
is reminiscent of the working principle of optical lattices;
moreover, very much in analogy to experiments performed with ultracold atoms in optical lattices,
the SAW setup offers similarly attractive features such as \textit{in-situ} tunable
system parameters and favourable scaling properties.
Furthermore, the applicability of the derived results is not limited to
electron traps but is more general;
in principle, all generalizations to quasiparticles with an internal level structure that
can be used to realize the model from Eq.~\eqref{eq:model} are candidates for
a realization of the proposed magnetic traps.
Quantitatively, the projected trap depths should allow for the implementation
of thermally robust and low-loss magnetic traps with state-of-the-art technology and
high $g$-factor materials such as InAs, InSb or dilute magnetic semiconductors.
With the possibility to reach yet unexplored parameter values,
especially in the low-temperature and strong-interaction regime
of the Fermi-Hubbard model,
solid-state magnetic lattices may constitute a novel platform for
studying superfluidity, quantum magnetism and strongly correlated
electrons in periodic systems.

Finally, we discuss possible future research directions.
(\textit{i}) By contrast with effectively one-dimensional systems discussed
in this work, two-dimensional lattices with vastly different geometries
might be studied.
Due to the flexibility of SAW-based setups, these lattice geometries
could be altered during an experiment.
By dynamically modulating the lattice, this may allow for the investigation
of intricate band structures or resonant coupling between different Bloch bands,
akin to experiments with shaken optical lattices \cite{gemelke05,lignier07,struck11,struck12}.
(\textit{ii}) Instead of considering electrons with two Zeeman-split
internal spin states, quasiparticles with a richer internal energy-level
structure might be examined (e.g., spin-$3/2$ holes).
Here, one interesting prospect could be the realization of tunable
subwavelength potential barriers for quasiparticles on the nanoscale,
in close analogy to dark-state optical lattices with subwavelength
spatial structure \cite{lacki16,wang18}.
(\textit{iii}) Apart from the two possible implementations studied in this
work, other implementations may be considered, either as stand-alone
alternatives or in combination with, e.g., SAWs.
Specifically, nanoengineered vortex arrays have been considered
in the past both for magnetic atom traps \cite{romero-isart13} and
strong magnetic modulations of Bloch electrons in 2DEGs \cite{movilla11}.
(\textit{iv}) Since we have only considered one-dimensional lattices,
anisotropies of system parameters were negligible so far.
In contrast, in two-dimensional systems, anisotropic effective electron masses or $g$-factors
can lead to strongly non-uniform potential landscapes and anisotropic
tunneling matrix elements.
Besides that, SOI can itself be a strongly anisotropic interaction,
thus modulating the SOI-induced hopping amplitude $t^\lambda_\pm$
($\lambda = \alpha_\mathrm{R}, \beta_\mathrm{D}$ in the presence of Rashba or Dresselhaus
SOI, respectively) in a way that it becomes anisotropic.
In this way, the effect of anisotropic hopping on the phase diagram
of a (spin-dependent) Fermi-Hubbard model might be studied, inheriting
its rich physics from a number of versatile material properties.

%% %%%%%%%%%%%%%%%%%%%%%%%%%%%%%%%%%%%%%%%%%%%%%%%%%%%%%%%%%%%%%%%%%%%%%

%% %%%%%%%%%%%%%%%%%%%%%%%%%%%%%%%%%%%%%%%%%%%%%%%%%%%%%%%%%%%%%%%%%%%%%
%% ACKNOWLEDGMENTS
\begin{acknowledgments}
\textit{Acknowledgments}.\textemdash 
J. K. and J. I. C. acknowledge support by the DFG within the Cluster of
Excellence NIM.
M. J. A. S. would like to thank the Humboldt foundation for financial support. 
G. G. acknowledges support by the Spanish Ministerio de Econom\'{i}a y
Competitividad through the Project No. FIS2014-55987-P and thanks MPQ for
hospitality.
Work at Harvard was supported by NSF, Center for Ultracold Atoms, CIQM,
Vannevar Bush Fellowship, AFOSR MURI and Max Planck Harvard Research Center
for Quantum Optics.
H. Huebl acknowledges support by the DFG Priority Programm SPP 1601 (HU1986/2-1).
This work was also partially funded by the European Union through the
European Research Council grant QUENOCOBA, ERC-2016-ADG (Grant No. 742102).
J. K. and M. J. A. S. thank Mihir Bhaskar, Ruffin Evans, Kristiaan de Greeve,
Hubert Krenner, Christian Nguyen, Lieven Vandersypen, Achim Wixforth, and
Peter Zoller for fruitful discussions.

$^*$J. K. and M. J. A. S. contributed equally to this work.
\end{acknowledgments}
%% %%%%%%%%%%%%%%%%%%%%%%%%%%%%%%%%%%%%%%%%%%%%%%%%%%%%%%%%%%%%%%%%%%%%%

%% %%%%%%%%%%%%%%%%%%%%%%%%%%%%%%%%%%%%%%%%%%%%%%%%%%%%%%%%%%%%%%%%%%%%%
%% %%%%%%%%%%%%%%%%%%%%%%%%%%%%%%%%%%%%%%%%%%%%%%%%%%%%%%%%%%%%%%%%%%%%%
%% BIBLIOGRAPHY

%% %%%%%%%%%%%%%%%%%%%%%%%%%%%%%%%%%%%%%%%%%%%%%%%%%%%%%%%%%%%%%%%%%%%%%

%% %%%%%%%%%%%%%%%%%%%%%%%%%%%%%%%%%%%%%%%%%%%%%%%%%%%%%%%%%%%%%%%%%%%%%
%% %%%%%%%%%%%%%%%%%%%%%%%%%%%%%%%%%%%%%%%%%%%%%%%%%%%%%%%%%%%%%%%%%%%%%
%% APPENDICES
\appendix

%% %%%%%%%%%%%%%%%%%%%%%%%%%%%%%%%%%%%%%%%%%%%%%%%%%%%%%%%%%%%%%%%%%%%%%
%% APPENDIX 1:
%% Magnus expansion
\section{Beyond the RWA \label{sec:beyond-rwa}}

\begin{figure}
\includegraphics[width=1\columnwidth]{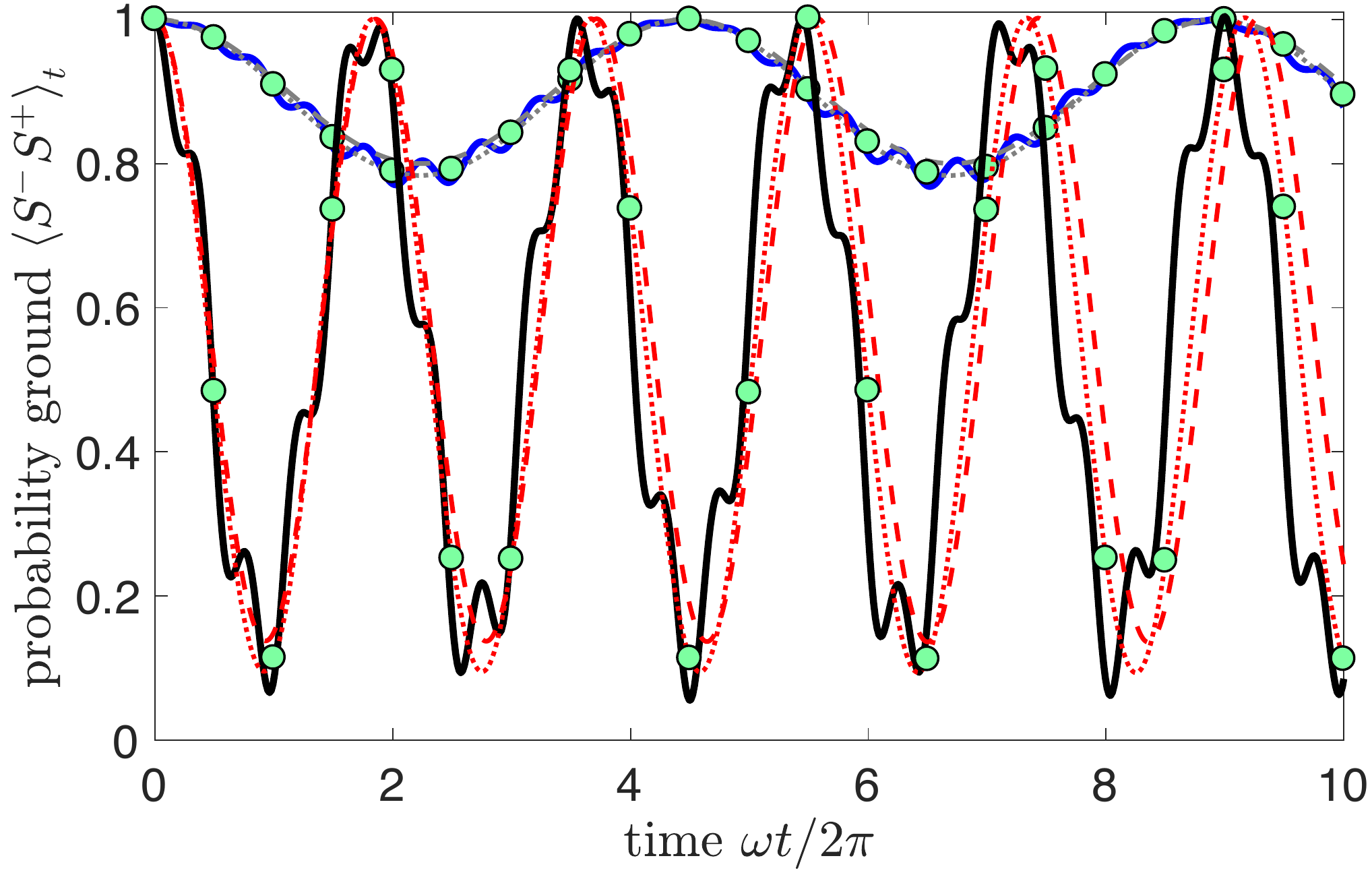}
\caption{\label{fig:magnus-expansion-vs-num-exact-propagation}(color online).
Numerical simulation of the dynamics generated by the time-dependent (i.e., without any RWA)
Hamiltonian (\ref{eq:model}) for $\Omega(\hat z)=\Omega_0=0.1\omega$ (blue solid line) and $\Omega_0=0.5\omega$ (black solid line), respectively. 
The corresponding dashed (dotted) lines refer to the dynamics generated by the time-independent zeroth-order (second-order) Floquet Hamiltonian $H_{F}$, 
with dots highlighting the results according to the second-order Floquet Hamiltonian $H_{F}$ at stroboscopic times $t_{n}=nT/2$.
The initial state has been set as $\left|\Psi\right>_{0}=\left|\downarrow\right>$. 
Other numerical parameters: $\Delta/\omega=0.2$.
}
\end{figure}

A fundamental limitation in the above discussion stems from the condition $\Omega_0 \ll \omega$ necessary
for the RWA to be justified. Due to this restriction, Rabi frequencies, and hence ultimately the trap depths,
are limited to values much smaller than the driving frequency $\omega$.
One way to lift this built-in restriction is to drop the RWA,
keeping counter-rotating terms $\propto \Omega(\hat z) \sigma^{\pm} e^{\pm 2i \omega t}$ in the Hamiltonian
Eq.~\eqref{eq:model} which can be written in a rotating frame as
\begin{equation}
H = \Delta \sigma^z + \frac{\Omega(\hat z)}{2} \sigma^x + \frac{\Omega(\hat z)}{2} \left ( \sigma^+ e^{2i\omega t} + \sigma^- e^{-2i\omega t} \right ).
\end{equation}
If we now consider the corresponding
time-evolution operator evaluated at stroboscopic times $t_n = t_0 + nT/2$ with $T=2\pi/\omega$,
\begin{equation}
U(t_n) = \mathcal{T}_{\leftarrow} \exp \left ( i \int_{t_0}^{t_n} \mathrm{d} \tau H(\tau) \right ),
\end{equation}
a Magnus expansion \cite{bukov15} up to second order in $1/\omega$ yields
\begin{equation}
U(t_n, t_0) = \exp \left ( -i H_F[t_0] nT/2 \right ),
\end{equation}
with the stroboscopic Floquet Hamiltonian $H_F$ given by
\begin{equation}
H_F = H_F^{(0)} + H_F^{(1)} + H_F^{(2)} + ...,
\end{equation}
with the three lowest-order contributions
\begin{eqnarray}
H_F^{(0)} & = & \frac{\Delta}{2} \sigma^z + \frac{\Omega(\hat z)}{2} \sigma^x, \\
H_F^{(1)} & = & \frac{\Omega(\hat z)}{16\omega} \left ( 2\Delta \sigma^x - \Omega(\hat z) \sigma^z \right ), \\
H_F^{(2)} & = & - \frac{\Omega(\hat z)}{64\omega^2} \left ( 4 \Delta^2  + \Omega^2(\hat z) \right ) \sigma^x.
\end{eqnarray}
Numerical results of the dynamics generated by the zeroth- and second-order
results are compared with the dynamics generated by the full time-dependent
Hamiltonian [the internal Hamiltonian $h$ in Eq.~(\ref{eq:model}), without RWA]
in Fig.~\ref{fig:magnus-expansion-vs-num-exact-propagation}.
From the numerical results we conclude that the (stroboscopic) characterization
of the system dynamics by $H_F$ works well only if $\Omega_0 \lesssim \omega$.
In this regime, even at higher orders we still obtain a time-independent
periodic Hamiltonian which allows for the implementation of magnetic
(super-)lattices.  

%% %%%%%%%%%%%%%%%%%%%%%%%%%%%%%%%%%%%%%%%%%%%%%%%%%%%%%%%%%%%%%%%%%%%%%
%% APPENDIX 2:
%% Non-adiabatic spin flips
\section{\label{sec:non-adiabatic-transitions}Spin-flip transitions in magnetic
traps and lattices}

Based on Ref.~\cite{sukumar97}, we investigate undesired spin-flip losses from a magnetic trap.
We consider the model
\begin{equation}
H=\frac{\hat p^{2}}{2m}+\omega_{0}\sigma^{z}+\Omega(\hat z)\cos(\omega t)\sigma^{x},
\end{equation}
which, in a rotating frame and within a rotating-wave approximation,
can be written as
\begin{equation}
H=\frac{\hat p^{2}}{2m}+\Delta \sigma^{z}+\frac{\Omega(\hat z)}{2}\sigma^{x}=\hat T+h(\hat z),\label{eq:model-RWA}
\end{equation}
where $\hat T=\hat p^{2}/(2m)$ and
\begin{equation}
h(\hat z)=\frac{1}{2}\left(\begin{array}{cc}
\Delta & \Omega(\hat z)\\
\Omega(\hat z) & -\Delta
\end{array}\right).
\end{equation}
We introduce a unitary operator $U(\hat z)=\exp(-i\frac{\theta(\hat z)}{2}\sigma^{y})$
acting on the internal states such that
\begin{eqnarray}
\vert+\rangle_{\theta} & = & U(\hat z)\vert\uparrow\rangle,\label{eq:unitary-trafo-pm}\\
\vert-\rangle_{\theta} & = & U(\hat z)\vert\downarrow\rangle.\nonumber 
\end{eqnarray}
Note that $U^{\dagger}(\hat z)$ rotates the
effective magnetic field to be parallel to the \textit{z} axis. The
transformed Hamiltonian $\tilde{H}$ takes the form
\begin{eqnarray}
\tilde{H} & = & U^{\dagger}(\hat z)HU(\hat z)\\
 & = & \hat T+\left[U^{\dagger}(\hat z)\hat TU(\hat z)-\hat T\right]+U^{\dagger}(\hat z)\left[h(\hat z)\right]U(\hat z)\label{eq:h-tilde}\nonumber\\
 & = & \hat T+\Delta T+\varepsilon(\hat z)\tilde \sigma^{z},\nonumber 
\end{eqnarray}
where $\Delta T=\left[U^{\dagger}(x)\hat TU(x)-\hat T\right]$, $\varepsilon(\hat z) = \frac{1}{2}\sqrt{\Delta^{2}+\Omega^{2}(\hat z)}$
and $\tilde \sigma^z = |+\rangle\langle+|-|-\rangle\langle-|$.
The adiabatic approximation amounts to neglecting the contribution
which stems from $\Delta T$ \cite{sukumar97}.
This is justified provided that $\chi = \omega_\mathrm{HO}/|\Delta| \ll 1$,
i.e., that the potentials defined by $\varepsilon$ and $-\varepsilon$
are sufficiently separated in energy.

%% %%%%%%%%%%%%%%%%%%%%%%%%%%%%%%%%%%%%%%%%%%%%%%%%%%%%%%%%%%%%%%%%%%%%%
%% APPENDIX 3:
%% Extended Hubbard model with additional spin-flip assisted hopping
\section{Spin-flip assisted tunneling processes in magnetic lattices \label{app:hubbard}}

In Eq.~\eqref{eq:hubbard2} in the main text, we present an extended
Hubbard model which includes both next-nearest (spin-conserving) neighbour
hopping ($\sim t_c$, compare with Eq.~\eqref{eq:hubbard}) and nearest neighbour
(spin-flip assisted) hopping ($\sim t_\pm$) processes.
In the following, we show how this Hamiltonian and, more specifically,
the additional hopping term $\sim t_\pm$ can be constructed with
the aid of additional RF driving fields.

Starting from Eq.~\eqref{eq:model}, we consider two auxiliary time-dependent
fields in addition to the field $\mathbf{B}(\mathbf{r},\omega t)$:
(\textit{i}) The driving field $\mathbf{B}_\text{dr}(t) = B_\text{dr} \cos(\omega_2 t) \mathbf{\hat x}$, a second rapidly oscillating
transverse field, is weaker than the RF field $\mathbf{B}_\perp(\mathbf{r},\omega t)$ which
provides the lattice and detuned from it so as to be resonant with the
energy difference between the two local spin directions.
(\textit{ii}) The third time-dependent field $\mathbf{B}_3 = B_3 \cos(\omega_3 t) \mathbf{\hat z}$
is slowly varying and parallel to the constant field $\mathbf{B}_{||}$ which provides the
Zeeman splitting; its purpose is to (partially) compensate
the longitudinal components that $\mathbf{B}_\text{dr}$ acquires in the adiabatic frame.

In the presence of these additional fields, two new terms appear in the
model of Eq.~\eqref{eq:model},
\begin{align}
  \label{eq:appHubbard-1}
  H_\text{dr} = \frac{\hat p^2}{2m} &+ \omega_0 \sigma^z +\Omega(\hat z)\cos(\omega t) \sigma^x \\
& {}+\nonumber
  \Omega_\text{dr}\cos(\omega_2t) \sigma^x + \Omega_3\cos(\omega_3t)\sigma^z,
\end{align}
where $\Omega_\text{dr}=\gamma B_\text{dr}$ and $\Omega_3=\gamma B_3$.
In the following, we require
$\omega,\omega_2\gg|\omega-\omega_2|\equiv\delta\approx\omega_3$ as well
as $|\Omega_0|\gg|\Omega_\text{dr}|,|\Omega_3|$.

Defining a rotating frame by $\ket{\psi^\text{rot}_t}=U_t\ket{\psi_t}$ (where
$\ket{\psi_t}$ denotes a solution of the Schr\"odinger equation in the lab frame)
with $U_t=\exp(it\omega \sigma^z)$, we obtain the Hamiltonian in the rotating
frame as
\begin{align}
  H^\text{rot}_\text{dr} = \frac{\hat p^2}{2m} &+ \Delta \sigma^z + 
\frac{\Omega(\hat z)}{2}\sigma^x
+ \frac{\Omega(\hat z)}{2}\left[ \ketbra{\uparrow}{\downarrow}e^{i2\omega t}+\mathrm{h.c.} \right] \nonumber\\
& 
+\frac{\Omega_{dr}}{2}\left[\ketbra{\uparrow}{\downarrow} ( e^{i\delta t}+e^{i(\omega+\omega_2)t}  ) + \mathrm{h.c.} \right ]\nonumber\\
& 
+ \Omega_3\cos(\omega_3t)\sigma^z.
\end{align}
Within a RWA, where we keep only the constant and slowly oscillating terms,
we obtain
\begin{align}
  H^\text{rot}_\text{dr} = \frac{\hat p^2}{2m} &+ \Delta \sigma^z + \frac{\Omega(\hat z)}{2}\sigma^x
+\frac{\Omega_{dr}}{2}\left[e^{i\delta t}\ketbra{\uparrow}{\downarrow}+\mathrm{h.c.} \right]\nonumber\\
& 
+ \Omega_3\cos(\omega_3t)\sigma^z.
\end{align}
Now, by employing the unitary transformation $U(\hat z)$ introduced in the main text,
we can (locally) diagonalize the constant contribution stemming from
$\hat p^2/(2m) + h_\text{RWA}(\hat z)$ [see Sec.~\ref{ssec:single-particle}].
Then, neglecting the non-adiabatic correction due to $\Delta T$ and simplifying
the resulting expressions yields
\begin{align}\label{eq:appHubbard-2}
\tilde{H} & = \frac{\hat p^2}{2m} + \varepsilon(\hat z) \tilde \sigma^z \nonumber\\
& + \left [\frac{\Omega_\text{dr}}{2} \cos^2 \vartheta \cos(\delta t) - 2 \Omega_3 \sin \vartheta \cos \vartheta \cos(\omega_3 t)
\right ] \tilde \sigma^x \nonumber\\
& + \left [ 2 \Omega_\text{dr} \sin \vartheta \cos \vartheta \cos(\delta t) + \Omega_3 (\cos^2 \vartheta - \sin^2 \vartheta ) \cos(\omega_3 t)
\right ] \tilde \sigma^z,
\end{align}
where $\vartheta := \theta(\hat z)/2 = \arcsin [ \frac{\Omega(\hat z)}{\sqrt{\Omega^2(\hat z) + \Delta^2}} ]/2$
and $\tilde \sigma^z = \ketbra{+}{+}-\ketbra{-}{-}$, $\tilde \sigma^x = \ketbra{+}{-}+\ketbra{-}{+}$.
Clearly, in comparison with Eq.~\eqref{eq:adiabatic-model}, we get additional
contributions due to the additional time-dependent fields.

We now use the fact that the newly introduced driving fields
are relatively weak compared to the fields considered in the main text
and treat these terms as a perturbation to the tight-binding model in Eq.~\eqref{eq:hubbard}.
Furthermore, from Eq.~\eqref{eq:appHubbard-2},
it becomes clear that the third driving field $\mathbf{B}_3$ can be used to
compensate for undesired (time-dependent) on-site terms due to $\mathbf{B}_\text{dr}$.
At the resonance $\omega_3 = \delta$ and within a rotating frame
$U^\text{rot2}_t = \exp(it\delta \tilde \sigma^z)$, the Hamiltonian \eqref{eq:appHubbard-2}
can be further simplified and a RWA with respect to $2\delta$ can be performed,
given that the off-resonant spin-flip terms oscillate much faster
than their strength.
Eventually, we obtain the extended Fermi-Hubbard model
\begin{eqnarray}\label{eq:appHubbard-3}
H_{\text{FH3}} & = & -t_c\sum_{\langle\langle i,j\rangle\rangle,s}(c_{is}^{\dagger}c_{js}+\text{h.c.})
-t_\pm\sum_{\langle i,j\rangle,s}(c_{is}^{\dagger}c_{j\bar s}+\text{h.c.}) \nonumber \\
 &  & + \sum_{i,s} \mu_{is} n_{is} + \sum_{s,s^\prime} \sum_{ijkl} U_{ijkl} c_{is^{\prime}}^{\dagger}c_{js}^{\dagger}c_{ls}c_{ks^{\prime}},
\end{eqnarray}
which reduces to Eq.~\eqref{eq:hubbard2} at the resonance $\delta = \Delta$.
Here, the nearest-neighbour tunneling is characterized by
$t_\pm=\langle w_j | \frac{\Omega_\text{dr}}{2} \cos^2\vartheta - 2\Omega_3 \sin \vartheta \cos \vartheta | w_{j+1} \rangle$
with the Wannier function $w_j$ located at lattice site $j$.

%% %%%%%%%%%%%%%%%%%%%%%%%%%%%%%%%%%%%%%%%%%%%%%%%%%%%%%%%%%%%%%%%%%%%%%
%% APPENDIX 4:
%% Implementation I: Superconducting circuit
\section{Implementation I: Superconducting circuit \label{app:ler}}

In the following, we describe the magnetic field due to an electric
current density $\textbf{J}$ by the Biot-Savart law. Since we are
dealing with AC fields, this description can only be approximately valid.
A more precise picture follows from the Jefimenkov equations \cite{jackson}:
\begin{eqnarray}
\textbf{B}_{\mathrm{AC}}(\textbf{r},t)=\frac{\mu_{0}}{4\pi}\int_{V}\mathrm{d}^{3}\textbf{r}^{\prime} \ 
\bigg (
\textbf{J}(\textbf{r}^{\prime},t_\mathrm{ret})\times\frac{\textbf{r}-\textbf{r}^{\prime}}{|\textbf{r}-\textbf{r}^{\prime}|^3} \nonumber \\
+ \frac{1}{c} \frac{\partial \textbf{J}(\textbf{r}^{\prime},t_\mathrm{ret})}{\partial t}
\times\frac{\textbf{r}-\textbf{r}^{\prime}}{|\textbf{r}-\textbf{r}^{\prime}|^2}
\bigg ). \label{eq:jefimenkov}
\end{eqnarray}
where the right-hand side of the equation is evaluated at the retarded time
$t_\mathrm{ret} = t - |\textbf{r}-\textbf{r}^\prime|/c$
and $c$ denotes the speed of light in the dielectric medium.
However, since the time-dependence of the current density
$\textbf{J}(\textbf{r}^{\prime},t) \sim \exp(i\omega t)$, the correction
term in Eq.~(\ref{eq:jefimenkov}) is expected to be of the order of
$|\textbf{r}-\textbf{r}^\prime| \omega/c \sim d \omega /c $
with the distance $d$ between meandering wire and 2DEG.
The wires are located above the surface at $x = 0$.
For typical distances
$d \sim (0.1-1)~\mu\mathrm{m}$ and frequencies $\omega \sim (1-100) \mathrm{GHz}$,
the correction term in Eq.~(\ref{eq:jefimenkov}) may be neglected and the
Biot-Savart law is recovered which then accurately describes the induced
magnetic field due the electric current density $\textbf{J}$,
\begin{equation}
\textbf{B}_{\mathrm{AC}}(\textbf{r},t)=\frac{\mu_{0}}{4\pi}\int_{V}\mathrm{d}^{3}\textbf{r}^{\prime} \ 
\textbf{J}(\textbf{r}^{\prime},t)\times\frac{\textbf{r}-\textbf{r}^{\prime}}{|\textbf{r}-\textbf{r}^{\prime}|^3}.
\end{equation}
\noindent In the following, we assume the spatial extension of the meandering wire
to exceed the relevant size of the 2DEG, i.e., the trapping region.
This assumption guarantees the absence of finite-size effects
at the turning points of the meandering wire, i.e., we model each parallel
line in the meandering wire as an infinite wire which induces a magnetic field
on its own.
Also, we neglect boundary effects from the border of the 2DEG.
In the case of an infinitely long wire which runs parallel to the
$y$ axis (cf.~Fig.~\ref{fig:LER-setup}), the Biot-Savart law simplifies
to \cite{jackson}
\begin{equation}
\textbf{B}_{\mathrm{AC}}(\textbf{r}=(\rho,\phi,y),t)=\frac{\mu_{0}I(t)}{2\pi\rho}\textbf{e}_{\phi},
\end{equation}
\noindent where $I(t)$ denotes the current in a single wire.
In the presence of many parallel wires (whose current flow alternates
between the $+y$ and $-y$ directions),
which is the situation that accurately describes the setup sketched
in Fig.~\ref{fig:LER-setup}, the magnetic field at point $\textbf{r}$ is given by
\begin{eqnarray}
\textbf{B}_{\mathrm{AC}}(\textbf{r}&=&(x,y,z)) = -\sum_{n}^{N}\frac{\mu_{0}\textbf{I}_{n}(t)}{2\pi}\times\frac{\textbf{r}_{n}}{\textbf{r}_{n}^{2}}\label{eq:savart-sum}\\
 & = & \frac{\mu_{0}I_0 \cos(\omega t)}{2\pi}\sum_{n}^{N}\frac{(-1)^{n}}{(z-na)^2+x^{2}}  % \left(z\textbf{e}_{x}+(x-na)\textbf{e}_{z}\right)\nonumber 
\begin{pmatrix} x \\ 0 \\ z-na \end{pmatrix}, \nonumber
\end{eqnarray}
\noindent with the center of the wires positioned at $x=0$ and
given a time-dependent current amplitude $I(t) = I_0 \cos(\omega t)$
in each wire and the position vectors $\mathbf{r}_n$ which denote the position
at which the field is evaluated relative to the $n$th wire.
An exemplary field distribution $\textbf{B}_{\mathrm{AC}}(\textbf{r},t=0)$
is shown in Fig.~\ref{fig:LER-setup}(b).
Due to the translational symmetry along the axis parallel to the
wires, Eq.~(\ref{eq:savart-sum}) enables us to write the spin Hamiltonian, in the presence of an
additional external magnetic field, as
\begin{eqnarray}
H&= & \gamma\textbf{B}_{\mathrm{AC}}(\mathbf{\hat r},t)\cdot\mathbf{\sigma}+\gamma B_{\mathrm{ext}} \sigma^{z}\label{eq:hamilt-LER}\\
&=  & \frac{\gamma\mu_{0}I_0}{2\pi}\sum_{n=1}^{N}\frac{(-1)^{n}x}{\hat z^{2}-2na\hat z+n^{2}a^{2}+x^{2}}\sigma^{x}\cos(\omega t)\nonumber\\
 & & +  \frac{\gamma\mu_{0}I_0}{2\pi}\sum_{m=1}^{N}\frac{(-1)^{m}(\hat z-ma)}{\hat z^{2}-2ma\hat z+m^{2}a^{2}+x^{2}}\sigma^{z}\cos(\omega t)\nonumber\\
 & & + \gamma B_{\mathrm{ext}} \sigma^{z}.\nonumber
\end{eqnarray}

The induced electric field due to a time-dependent magnetic field
is described by Faraday's law, $\nabla\times\textbf{E}=-\partial\textbf{B}/\partial t$.
By (anti-)symmetries of the straight long wire and its magnetic field \textemdash{}
translations along the $y$ axis, rotations about $y$ axis, and the reflection
$y\rightarrow-y$ \textemdash the induced electric field points in
a direction parallel to the wire, i.e., along $y$. Hence, the induced
electric field should not affect the magnetic lattice along $z$.
The motional DOF along $y$ could experimentally be frozen out, e.g., via
the implementation of an etched channel.

We define $\omega_0 = g_\text{s} \mu_\mathrm{B} B_{\mathrm{ext}}$ and rewrite
(\ref{eq:hamilt-LER}) as

\begin{equation}
H = \left [ \omega_0 + \Omega_0^z (\hat z) \cos(\omega t) \right ] \sigma^{z} + \Omega_0^x (\hat z) \cos(\omega t) \sigma^x \label{eq:hamilt-2-LER}
\end{equation}

Next, we take a closer look at the spatial profiles of the Rabi frequencies
$\Omega_0^z(\hat z)$ and $\Omega_0^x(\hat z)$ in Eq.~\eqref{eq:hamilt-2-LER}.
The time-dependent field amplitudes in Eq.~\eqref{eq:hamilt-2-LER} can be exactly
expressed via the Digamma function $\digamma$ (logarithmic derivative
of the $\Gamma$ function; \cite{gradshteyn00,weisstein}). 
Denoting the two sums appearing there as $b_x$ and $b_z$,
respectively,  setting $a=1$ and using $\xi=-z+ix$, it holds that 
\begin{align}
b_z+ib_x = &-\frac{1}{2} \digamma\left(\xi/2+\lfloor
    (N-1)/2\rfloor+1\right)+\frac{1}{2} \digamma(\xi/2)\nonumber\\
    &+ \frac{1}{2}\digamma\left([\xi+1]/2 +\lfloor N/2\rfloor
 \right)-\frac{1}{2}\digamma([\xi+1]/2)\nonumber\\
\stackrel{N\to\infty}{=}&
\frac{1}{2}\left(\digamma(\xi/2)-\digamma([\xi+1]/2)\right). 
\end{align}
For $N\gg z\gg1$ the real and imaginary parts of this function are
(approximately) periodic with period $1$ and have zeros at integer
(half-integer) values of $z$, respectively. 
For an odd number of
wires, the $z$ ($x$) field components are antisymmetric (symmetric) with respect
to the axis $z=z_s\equiv(N-1)/2$; (for even $N$,  $B_z$ is symmetric and $B_x$
antisymmetric).
The fields are well approximated by $b_z+ib_x\propto\exp(-i\pi
z)$, with errors less than $0.1\%$ but not approaching zero as $N\gg
z\to\infty$.
Using properties of the Digamma function, we can write
\begin{align}
  b_z+ib_x =& 
  \frac{1}{2}\sum_{l=0}^{\lfloor N/2\rfloor-1}  \frac{1}{l+(\xi+1)/2}\\
&- \frac{1}{2}\sum_{l=0}^{\lfloor (N-1)/2\rfloor}  \frac{1}{l+\xi/2}.
\nonumber
\end{align}

\begin{center}
\begin{figure}[t!]
\includegraphics[width=9cm]{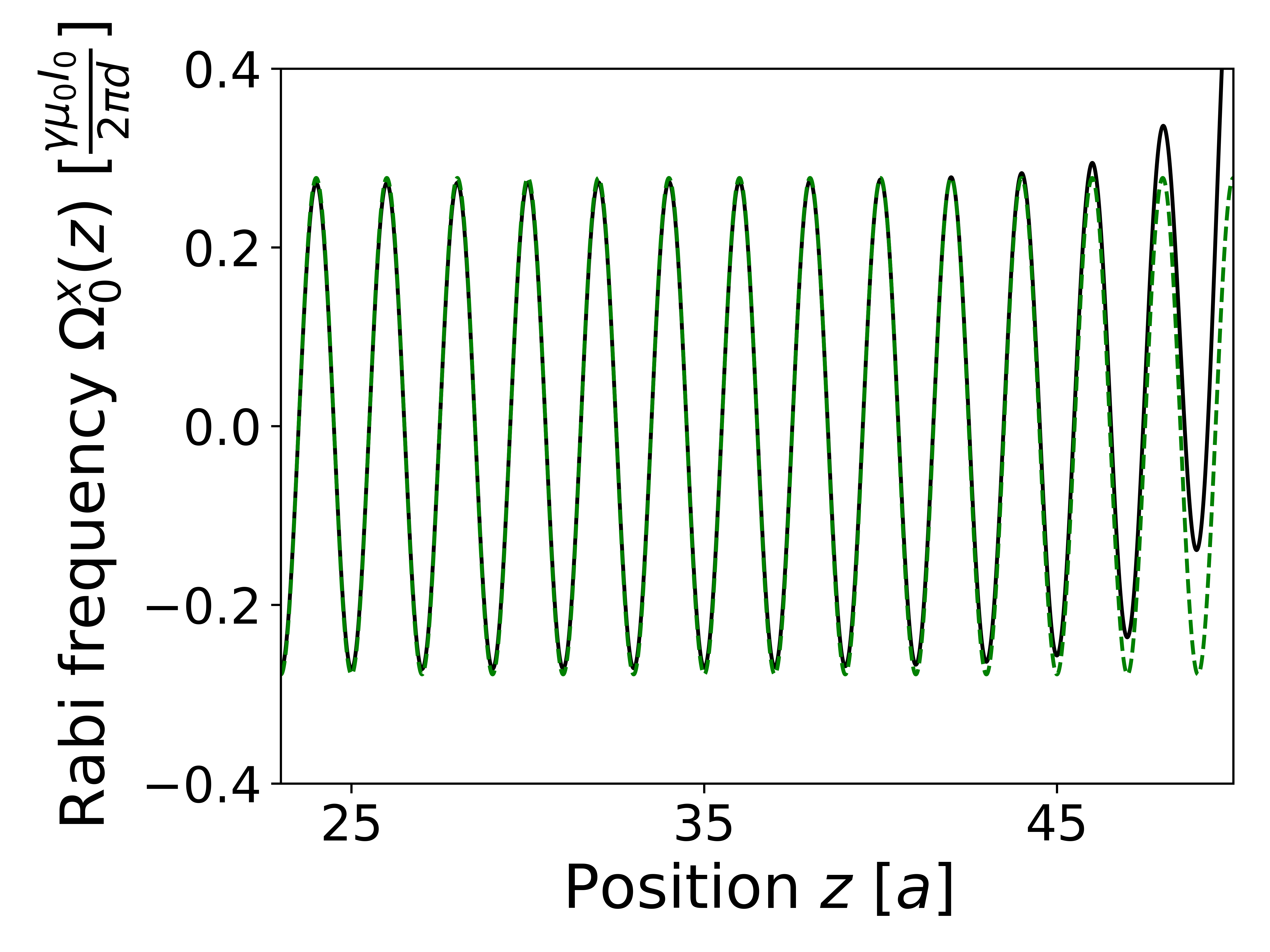}
\caption{\label{fig:bx-cut}(color online). Spatial pattern of Rabi frequency
(at given time), compare $\Lambda(\textbf{x})$ in Sec.~\ref{sec:theory}.
\textit{Black (solid):} calculated from Eq.~(\ref{eq:hamilt-LER}), \textit{green (dashed):}
$\sin$-fit.
At the ends of the meandering wire, i.e. at the edges of the lattice, finite-size
effects become apparent, but in the center of the lattice $\Lambda(\hat z)$ is
well-described by the sinusoidal fitting curve.
Parameters: $d = a$ and $N=50$ wires.}
\end{figure}
\end{center}

As shown in Fig.~\ref{fig:bx-cut}, the spatial dependence of $\Omega_0^x (\hat z)$
and $\Omega_0^z (x)$ (not shown) can (depending on the choice of parameters)
be well-described by a \textit{sine} function.
Hence, we can approximately write
\begin{eqnarray}
H = \left [ \omega_0 + \Omega_0^z \sin(\frac{\pi}{a} \hat z + \varphi) \cos(\omega t) \right ] & \sigma^{z} \label{eq:hamilt-3-LER} \\
+ \Omega_0^x \ \sin(\frac{\pi}{a} \hat z) \cos(\omega t) & \sigma^x, \nonumber
\end{eqnarray}
\noindent where $\varphi$ denotes a phase shift between $\Omega_0^x(\hat z)$ and $\Omega_0^z(\hat z)$.
%**********************************************************************%
%**********************************************************************%
In the center region, where finite-size effects are negligible, the Rabi
frequencies $\Omega_0^x$ and $\Omega_0^z$ are approximately given by
\begin{eqnarray}
\Omega_{0}^z & = & \gamma \frac{\mu_{0}I_0}{\pi a}\sum_{n=0,1,..}\frac{ (-1)^n \left ( n + 1/2 \right ) }{(n+1/2)^2 + (d/a)^2},\label{eq:rabi-x}\\
\Omega_{0}^x & = & \gamma \frac{d}{a} \frac{\mu_{0}I_0}{\pi a}\sum_{n=0,1,..}\frac{ (-1)^n }{(n+1/2)^2 + (d/a)^2}.\label{eq:rabi-z}
\end{eqnarray}
\noindent The expressions (\ref{eq:rabi-x}) and (\ref{eq:rabi-z}) become
exact in the limit of infinitely many wires, $N \rightarrow \infty$.
For all practical purposes considered in this work, $\Omega_0^z$ is very
small such that $\Omega_0^z \ll \omega_0$ and it may be safely neglected.

%% %%%%%%%%%%%%%%%%%%%%%%%%%%%%%%%%%%%%%%%%%%%%%%%%%%%%%%%%%%%%%%%%%%%%%
%% APPENDIX 5:
%% Implementation II: Surface acoustic waves
\section{Implementation II: Surface acoustic waves \label{app:SAWs}}

\subsection{Magnetization dynamics and effective magnetic field}
%%%%%%%%%%%%%%%%%%%%%%%%%%%%%% Constitutive relations
\textit{Constitutive relations for magnetoelastic couplings.}\textemdash
The governing constitutive relations for magnetostriction \cite{robillard09} read
\begin{eqnarray}
T_{ij} & = & c_{ijkl} u_{kl} - h_{kij} H_k, \label{eq:constitutive-stress} \\
B_{\text{dr},i} & = & h_{ijk} u_{ik} + \mu_{ij} H_j, \label{eq:constitutive-induction}
\end{eqnarray}
\noindent where $\underline{\underline{T}}$, $\mathbf{B}_\text{dr}$, $\mathbf{H}$ and $\underline{\underline{h}}$
denote the stress tensor, the magnetic induction,
the magnetic field (intensity vector) generated by a magnetoelastic wave and the effective piezomagnetic
tensor, respectively.
$\mu$ is the magnetic permeability and the strain field is defined as
$u_{kl}(\textbf{x}) = \left ( \partial u_k / \partial x_l + \partial u_l / \partial x_k \right ) / 2$.

Given Eq.~(\ref{eq:constitutive-induction}), we provide an estimate for
the effective driving field in the ferromagnet,
\begin{equation}
B_{\text{dr},1} \approx h k U = 2\pi h \frac{U}{\lambda}, \label{eq:b1-estimate}
\end{equation}
\noindent where $h$ denotes the magnetoelastic constant, $k$ is the wavevector and $U$ denotes
the amplitude of the displacement field.
For small strain-field amplitudes $kU \approx 10^{-6}$ and a magnetoelastic
constant $h=10$~T, this magnitude can be estimated as $B_{\text{dr},1} \approx 25~\mu$T \cite{dreher12}.

At ferromagnetic resonance, the effective magnetic field can be significantly
enhanced.
The response of a ferromagnet to small time-varying magnetic fields can
be described with the aid of Eq.~\eqref{eq:LLG}.
The resulting dynamical component of the magnetization $\mathbf{m}$
is given by
\begin{equation}
\mu_0|\mathbf{m}_\text{s}|\mathbf{m}=\bar \chi \mathbf{B}_{\text{dr}},
\end{equation}
where $\bar \chi$ denotes the Polder susceptibility which describes the
magnetic response of a ferromagnet to small time-varying magnetic fields
perpendicular to the magnetization equilibrium direction \cite{dreher12}.
In practical terms this means that the resulting effective magentic field
can be enhanced by about two orders of magnitude.

In a next step, the field at the 2DEG is then calculated from the field
distribution at the ferromagnetic thin film by discretizing the field
distribution at the film and summing up the dipole fields of these
volume elements.
At high strain amplitudes $kU \sim 10^{-4}-10^{-3}$ and a magnetoelastic
constant $h=(10-25)$~T, the relevant magnitude of the field at the 2DEG
can be numerically estimated as $B_1 \sim (10-100)$~mT.
In our numerical calculations, the amplitude of the displacement field,
the magnetoelastic coupling constant and the wavevector are input
parameters which determine the microwave field strength at the ferromagnetic
layer.

\subsection{Strain-induced potentials}

Starting from Eq.~\eqref{eq:hamilt-AL-and-ML} and in a suitable rotating frame, we obtain
\begin{eqnarray}
H_\mathrm{hyb}^\mathrm{rot}&=&\frac{\hat p^2}{2m} + V_\mathrm{SAW} \cos(k \hat z) \cos(\omega t) \\
&&+\frac{\omega_0}{2}\sigma^z + \frac{\Omega(\hat z)}{2} \left ( \sigma^x + e^{2i\omega t}\sigma^+ + e^{-2i\omega t}\sigma^- \right ),\nonumber
\end{eqnarray}
with $\sigma^+=\ket{\uparrow}\bra{\downarrow}$ and $\sigma^-=\ket{\downarrow}\bra{\uparrow}$.
Following the procedure outlined in Refs.~\cite{rahav03,rahav03b} and using
results from \cite{schuetz17}, we derive an effective time-independent
Hamiltonian up to second order in $1/\omega$ which reads
\begin{equation}
H_\mathrm{hyb}^\mathrm{eff}=\frac{\hat p^2}{2m} + \tilde \varepsilon(\hat z) \tilde \sigma^z
+ \left ( \frac{q^2}{8}E_\mathrm{S} + \frac{r}{4}|\Delta| \right )\sin^2(k \hat z),
\end{equation}
with $\tilde \varepsilon(\hat z) = \frac{1}{2}\sqrt{\Omega^2(\hat z) + \tilde \Delta^2}$,
$\tilde \Delta = |\Delta| + \Omega_0^2/(8E_\mathrm{S})$, $q=V_\text{SAW}/E_\text{S}$
and $r=\Omega_{0}^{2}/(4E_{\text{S}}\Delta)$.
For typical parameter values $r\ll 1$, $q^2/8\ll1$ and $\Omega_0\ll |\Delta|$, we
obtain the simplified form
\begin{equation}\label{eq:hybrid-effective-pot}
H_\mathrm{hyb}^\text{eff}\approx\frac{\hat p^2}{2m} + \frac{|\Delta|}{2} \tilde \sigma^z +
\left [ \frac{V_\text{SAW}^2}{8E_\mathrm{S}} - \frac{\Omega_0^2}{4|\Delta|} \tilde \sigma^z \right ] \sin^2(k \hat z),
\end{equation}
which coincides with the result given in Eq.~\eqref{eq:hamilt-AL-and-ML-effective}.
Writing Eq.~\eqref{eq:hybrid-effective-pot} in the form
$H_\mathrm{hyb}^\text{eff} = \hat p^2/2m + |\Delta|/2 \tilde \sigma^z + V_\mathrm{hyb} \sin^2(k\hat z)$,
we find that the spin-dependent potential amplitudes read
%% %%%%%%%%%%%%%%%%%%%%%%%% EQUATION 9 %%%%%%%%%%%%%%%%%%%%%%%%%%%%%%%%%%%
\begin{eqnarray}\label{eq:hybrid-AL-and-ML-amplitudes}
\braket{+|V_\mathrm{hyb}|+} & \approx & \frac{\Omega_0^2}{4|\Delta|} - \frac{q^2}{8}E_\mathrm{S}, \nonumber \\
\braket{-|V_\mathrm{hyb}|-} & \approx & - \frac{\Omega_0^2}{4|\Delta|} - \frac{q^2}{8}E_\mathrm{S}.
\end{eqnarray}
%% %%%%%%%%%%%%%%%%%%%%%%%% EQUATION 9 %%%%%%%%%%%%%%%%%%%%%%%%%%%%%%%%%%%
The resulting trap depths are depicted in Fig.~\ref{fig:hybrid-potential}.

\subsection{Stability analysis of hybrid magnetic and strain-induced traps \label{app:gsa}}

The discussion in this section completes the discussion of hybrid
magnetic and strain-induced traps and is devoted to the stability analysis
of such traps, meaning whether or not electrons can be trapped in
time-dependent trapping potentials of the kind of those featured in
Eq.~\eqref{eq:hamilt-AL-and-ML}.
%% %%%%%%%%%%%%%%%%%%%%%%%% FIGURE 5 %%%%%%%%%%%%%%%%%%%%%%%%%%%%%%%%%%%
\begin{figure}[t!]
  \centering
   \includegraphics[width=0.5\textwidth]{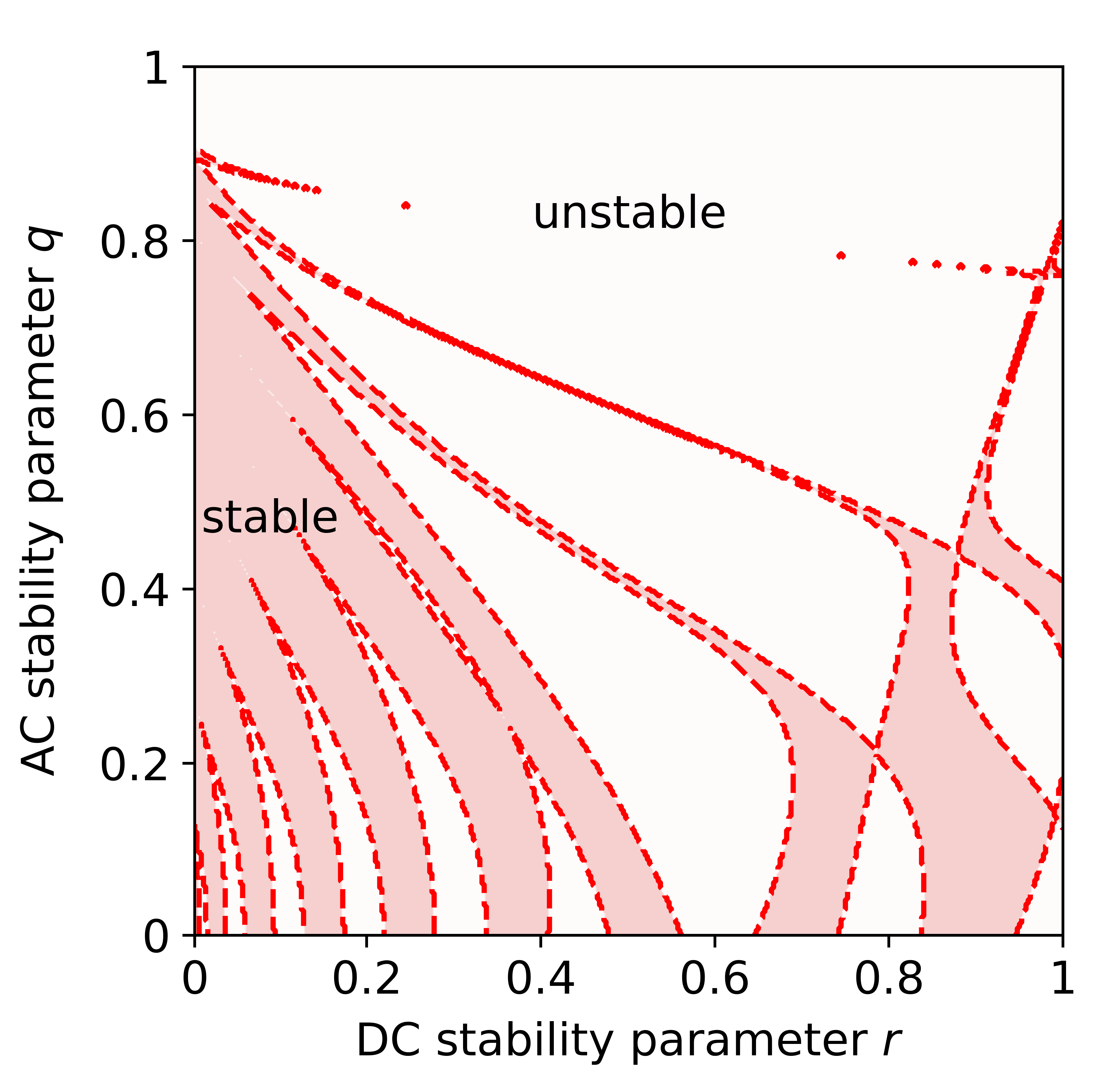}
\caption{
Stability diagram of Eq.~(\ref{eq:generalized-mathieu}) with stability
paramters $q = V_\text{SAW}/E_\text{S}$ and $r=\Omega_{0}^{2}/(4E_{\text{S}}\Delta)$.
Red areas denote regions of stable trapping, i.e.~stable solutions
of Eq.~(\ref{eq:generalized-mathieu}), and white areas, in turn, denote
unstable areas.
On the $r = 0$ axis, the standard Mathieu equation is recovered which,
for a purely time-dependent drive, yields stable trajectories in the
region $0 \leq q \lesssim 0.908$.
Other numerical parameters: $\eta=0.1$.
}
\label{fig:gsa-diagram}
\end{figure}
%% %%%%%%%%%%%%%%%%%%%%%%%% FIGURE 5 %%%%%%%%%%%%%%%%%%%%%%%%%%%%%%%%%%%

Starting from Eq.~\eqref{eq:hamilt-AL-and-ML}, we would like to predict
whether a given set of
parameters $\{ m, \ \omega, \ \omega_0, \ V_\text{SAW}, \ \Omega_0 \}$
gives rise to a stable (hybrid strain-induced and magnetic) trap or not.
To this end, we first derive the coupled Heisenberg equations of
motion for the set of observables $\{ \braket{z}, \braket{p}, \braket{\sigma^x},
\braket{\sigma^y}, \braket{\sigma^z} \}$ within a RWA.

\textit{Equations of motion.}\textemdash
In order to determine the EOMs of interest, we consider the
time evolution ($\tau = \omega t/2$) of the operators
$\tilde z:=k\hat z, \tilde p := \mathrm{d}\tilde z/\mathrm{d} \tau, \sigma^x, \sigma^y, \sigma^z$
which is given by the Heisenberg EOMs,
\begin{eqnarray*}
\langle \dot{\tilde z} \rangle & = & \langle \tilde p \rangle, \label{eq:heisenberg-x-2} \\
\langle \dot{\tilde p} \rangle & = & \frac{2\Omega_0}{E_S} \langle \sin(\tilde z) \rangle
\cos (2\tau) + \frac{V_\text{SAW}}{2 E_S} \langle \sin(\tilde z) \rangle \langle \sigma^x \rangle,
\label{eq:heisenberg-p-2} \\
\langle \dot \sigma^x \rangle & = & -2\frac{\Delta}{\omega} \langle \sigma^y \rangle, \\
\langle \dot \sigma^y \rangle & = & 2\frac{\Delta}{\omega} \langle \sigma^x \rangle -
\frac{V_\text{SAW}}{\omega} \langle \cos(\tilde z) \rangle \langle \sigma^z \rangle, \\
\langle \dot \sigma^z \rangle & = & \frac{V_\text{SAW}}{\omega}
\langle \cos(\tilde z) \rangle \langle \sigma^y \rangle, \label{eq:heisenberg-sigma-z-2}
\end{eqnarray*}
with $E_\mathrm{S} = m (\omega/k)^2/2$ and assuming that there exists no
significant correlation between external and internal DOFs,
i.e., decorrelated expressions such as,
e.g., $\langle \sin(\tilde z +\varphi) \sigma^i\rangle \approx \langle \sin(\tilde z +\varphi) \rangle \langle \sigma^i \rangle$.

\textit{Two limiting cases.}\textemdash
We consider the two limiting cases (\textit{i}) $\Omega_0 = 0$
and (\textit{ii}) $V_\text{SAW} = 0$:
(\textit{i}) At $\Omega_0 = 0$, we recover a Hamiltonian which is discussed
in great detail in Ref.~\cite{schuetz17}; in the limit $\tilde z \ll 1$,
the Heisenberg EOMs yield a Mathieu equation \cite{leibfried03} whose
stability diagram in terms of $V_\text{SAW}$ and
$E_\text{S} = m v_s^2/2$ is well-known, where $v_\text{s}$
denotes the speed of sound.
(\textit{ii}) For $V_\text{SAW} = 0$ and in the large-detuning regime $\Omega_0 \ll |\Delta|$,
an EOM can be derived which corresponds for a given spin state to a Hamiltonian of the form
$H = \hat p^2/(2m) + \Omega_0^2/(4|\Delta|) \sin^2(k \hat z)$.
Intuitively, these results agree very well with our expectation, since the
case (\textit{i}) coincides with a result known from the physics of trapped
ions; this is not surprising since only the electric field contributes.
On the other hand, case (\textit{ii}) reproduces an effective Hamiltonian
which is very familiar from optical lattices for cold (neutral) atoms
\cite{lewenstein07};
this finding, in turn, underlines the close relation between the proposed
magnetic traps and optical dipole traps which are both based on the AC Stark
effect.
In general, i.e., $V_\text{SAW}, \Omega_0 \neq 0$, the EOM leads to more
involved dynamics.
By adiabatic elimination of the internal DOFs, we obtain [corresponding to
the constructive case in Eq.~\eqref{eq:hybrid-AL-and-ML-amplitudes}]
an EOM of the form
%% %%%%%%%%%%%%%%%%%%%%%%%% EQUATION 9 %%%%%%%%%%%%%%%%%%%%%%%%%%%%%%%%%%%
\begin{equation}\label{eq:generalized-mathieu}
\ddot{\tilde z}+[r+2q\cos(2\tau)-r\cos(2\eta\tau)]\tilde z=0,
\end{equation}
%% %%%%%%%%%%%%%%%%%%%%%%%% EQUATION 9 %%%%%%%%%%%%%%%%%%%%%%%%%%%%%%%%%%%
with stability parameters $r=\Omega_{0}^{2}/(4E_{\text{S}}|\Delta|)$ and
$q=V_\text{SAW}/E_{\text{S}}$ and dimensionless quantities $\tilde x = kx$
and $\tau = \omega t/2$.
The ratio $\eta = |\Delta|/\omega$ is typically small in the RWA regime.
Based on Eq.~(\ref{eq:generalized-mathieu}), we extract stability
diagrams (to predict the stability of solutions to Eq.~(\ref{eq:generalized-mathieu}))
in terms of $q, \ r$ and $\eta$.
These diagrams can have an intricate structure,
see also Ref.~\cite{trypogeorgos16,leefer17}.
Here, we are mainly interested in the prediction of parameter
constellations that give rise to stable solutions of Eq.~(\ref{eq:generalized-mathieu}).
A prototypical stability diagram is shown in Fig.~\ref{fig:gsa-diagram}
for $\eta = 0.1$.
It can be seen that a $r=0$ cut in Fig.~\ref{fig:gsa-diagram} reproduces
the well-known result that stable behaviour of solutions to the Mathieu
equation occurs at $0 < q \lesssim 0.908$ for $r = 0$.
At $r > 0$, the stability properties can be rather sensitive
to slight changes in $q$.
An operation in the stable regime therefore requires a balanced choice
of these parameters.
However, Fig.~\ref{fig:gsa-diagram} shows that several values
$r > 0$ support a range of stable values $q$ which indicates that
operation in a stable regime is possible for a significant range of parameters.
Moreover, the numerical parameters used in
Fig.~\ref{fig:setup-SAWs} give rise to $q \ll 1$ which allows
for stable trajectories for many different $r$.
We conclude that, even in the presence of induced electric fields,
stable magnetic traps can be operated.


\begin{thebibliography}{10}
\bibitem{lewenstein07}M. Lewenstein, A. Sanpera, V. Ahufinger, B. Damski, A. Sen De, U. Sen,
\textit{Ultracold atomic gases in optical lattices: mimicking condensed matter physics and beyond},
Adv. Phys. \textbf{56}, 243 (2007).
\bibitem{bloch12}I. Bloch, J. Dalibard, and S. Nascimbene,
\textit{Quantum simulations with ultracold quantum gases},
Nat. Phys. \textbf{8}, 267 (2012).
\bibitem{bloch08}I. Bloch, J. Dalibard, W. Zwerger,
\textit{Many-body physics with ultracold gases},
Rev. Mod. Phys. \textbf{80}, 885 (2008).
\bibitem{barthelemy13}P. Barthelemy, and L. M. K. Vandersypen,
\textit{Quantum Dot Systems: a versatile platform for quantum simulations},
Ann. Phys. \textbf{525}, 808 (2013).
\bibitem{lee01}H. Lee, J. A. Johnson, M. Y. He, J. S. Speck, and P. M. Petroff,
\textit{Strain-engineered self-assembled semiconductor quantum dot lattices},
Appl. Phys. Lett. \textbf{78}, 105 (2001).
\bibitem{hanson07}R. Hanson, L. P. Kouwenhoven, J. R. Petta, S. Tarucha, and
L. M. K. Vandersypen,
\textit{Spins in few-electron quantum dots},
Rev. Mod. Phys. \textbf{79}, 1217 (2007).
\bibitem{alloing13}M. Alloing, A. Lemaitre, E. Galopin, and F. Dubin,
\textit{Optically programmable excitonic traps},
Sci. Rep. \textbf{3}, 1578 (2013).
\bibitem{schuetz10}M. J. A. Schuetz, M. G. Moore and C. Piermarocchi,
\textit{Trionic optical potential for electrons in semiconductors},
Nature Phys. \textbf{6}, 919 (2010).
\bibitem{rocke97}C. Rocke, S. Zimmermann, A. Wixforth, J. P. Kotthaus, G. B\"ohm, and G. Weimann,
\textit{Acoustically Driven Storage of Light in a Quantum Well},
Phys. Rev. Lett. \textbf{78}, 4099 (1997).
\bibitem{zimmermann99}S. Zimmermann, A. Wixforth, J. P. Kotthaus, W. Wegscheider, M. Bichler,
\textit{A Semiconductor-Based Photonic Memory Cell},
Science \textbf{283}, 1292 (1999).
\bibitem{ford17}C. J. B. Ford,
\textit{Transporting and manipulating single electrons in surface-acoustic-wave minima},
Phys. Status Solidi B \textbf{254}, 1600658 (2017).
\bibitem{folman02}R. Folman, P. Kr\"uger, J, Schmiedmayer, J. Denschlag, and Carsten Henkel,
\textit{Microscopic atom optics: from wires to an atom chip},
Adv. in At. Mol. Opt. Physics \textbf{48}, 263 (2002).
\bibitem{keil16}M. Keil, O. Amit, S. Zhou, D. Groswasser, Y. Japha, and R. Folman,
\textit{Fifteen Years of Cold Matter on the Atom Chip: Promise, Realizations, and Prospects},
J. Mod. Opt. \textbf{63}, 1840 (2016).
\bibitem{andre06}A. Andr\'{e}, D. Demille, J. M. Doyle, M. D. Lukin, S. E. Maxwell, P. Rabl, R. J. Schoelkopf, and P. Zoller,
\textit{A coherent all-electrical interface between polar molecules and mesoscopic superconducting resonators},
Nat. Phys. \textbf{2}, 636 (2006).
\bibitem{hou17}S. Hou, B. Wei, L. Deng, and J. Yin,
\textit{Chip-based microtrap arrays for cold polar molecules},
Phys. Rev. A \textbf{96}, 063416 (2017).
\bibitem{hensgens17}T. Hensgens, T. Fujita, L. Janssen, Xiao Li,
C. J. Van Diepen, C. Reichl, W. Wegscheider, S. Das Sarma, and L.
M. K. Vandersypen,
\textit{Quantum simulation of a Fermi-Hubbard model using a semiconductor quantum dot array},
Nature \textbf{548}, 73 (2017).
\bibitem{schuetz17}M. J. A. Schuetz, J. Kn\"orzer, G. Giedke, L. M. K. Vandersypen, M. D. Lukin, and J. I. Cirac,
\textit{Acoustic Traps and Lattices for Electrons in Semiconductors},
Phys. Rev. X \textbf{7}, 041019 (2017).
\bibitem{romero-isart13}O. Romero-Isart, C. Navau, A. Sanchez, P. Zoller, and J. I. Cirac,
\textit{Superconducting Vortex Lattices for Ultracold Atoms},
Phys. Rev. Lett. \textbf{111}, 145304 (2013).
\bibitem{tejada17}J. Tejada, E. M. Chudnovsky, R. Zarzuela, N. Statuto, J. Calvo-de la Rosa, P. V. Santos, and A. Hern\'{a}ndez-M\'{i}nguez,
\textit{Switching of magnetic moments of nanoparticles by surface acoustic waves}.
Europhys. Lett. \textbf{118}, 37005 (2017).
\bibitem{salasyuk17}
A. S. Salasyuk, A. V. Rudkovskaya, A. P. Danilov, B. A. Glavin, S. M. Kukhtaruk, M. Wang, A. W. Rushforth,
P. A. Nekludova, S. V. Sokolov, A. A. Elistratov,
D. R. Yakovlev, M. Bayer, A. V. Akimov, and A. V. Scherbakov,
\textit{Generation of a localized microwave magnetic field by coherent phonons in a ferromagnetic nanograting},
Phys. Rev. B \textbf{97}, 060404(R) (2018).
\bibitem{grimm00}R. Grimm, M. Weidem\"uller, Y. B. Ovchinnikov,
\textit{Optical dipole traps for neutral atoms},
Adv. Atom. Mol. Opt. Phys. \textbf{42}, 95 (2000).
\bibitem{redlinski05}P. Redlinski, T. Wojtowicz, T. G. Rappoport, A. Libal, J. K. Furdyna, and B. Janko,
\textit{Zero- and one-dimensional magnetic traps for quasiparticles in diluted magnetic semiconductors},
Phys. Rev. B. \textbf{72}, 085209 (2005).
\bibitem{berciu05}M. Berciu, T. G. Rappoport, and B. Janko,
\textit{Manipulating spin and charge in magnetic semiconductors using superconducting vortices},
Nature \textbf{435}, 71 (2005).
\bibitem{christianen98}P. C. M. Christianen, F. Piazza, J. G. S. Lok, J. C. Maan, W. van der Vleuten,
\textit{Magnetic traps for excitons},
Physica B \textbf{249}, 624 (1998).
\bibitem{murayama06}A. Murayama, M. Sakuma,
\textit{Nanoscale magnet for semiconductor spintronics},
Appl. Phys. Lett. \textbf{88}, 122504 (2006).
\bibitem{furdyna88}J. K. Furdyna,
\textit{Diluted magnetic semiconductors},
J. Appl. Phys. \textbf{64}, R29 (1998).
\bibitem{halm08}S. Halm, P. E. Hohage, J. Nannen, E. Neshataeva, L. Schneider, G. Bacher, Y. H. Fan, J. Puls, and F. Henneberger,
\textit{Manipulation of spin states in a semiconductor by microscale magnets},
J. Phys. D: Appl. Phys. \textbf{41}, 164007 (2008).
\bibitem{chen08}Y. S. Chen, S. Halm, E. Neshataeva, T. K\"ummell, G. Bacher, M. Wiater, T. Wojtowicz, and G. Karczewski,
\textit{Local control of spin polarization in a semiconductor by microscale current loops},
Appl. Phys. Lett. \textbf{93}, 141902 (2008).
\bibitem{jaksch98}D. Jaksch, C. Bruder, J. I. Cirac, C. W. Gardiner, and P. Zoller,
\textit{Cold bosonic atoms in optical lattices},
Phys. Rev. Lett. \textbf{81}, 3108 (1998).
\bibitem{ruostekoski02}J. Ruostekoski, G V. Dunne, and J. Javanainen,
\textit{Particle Number Fractionalization of an Atomic Fermi-Dirac Gas in an Optical Lattice},
Phys. Rev. Lett. \textbf{88}, 180401 (2002).
\bibitem{jaksch03}D. Jaksch, and P. Zoller,
\textit{Creation of effective magnetic fields in optical lattices: the Hofstadter butterfly for cold neutral atoms},
New J. Phys. \textbf{5}, 56 (2003).
\bibitem{dalibard11}J. Dalibard, F. Gerbier, G. Juzeli\={u}nas, and P. \"Ohberg,
\textit{Colloquium: Artificial gauge potentials for neutral atoms},
Rev. Mod. Phys. \textbf{83}, 1523 (2011).
\bibitem{goldman14}N. Goldman,G. Juzeli\={u}nas, P. \"Ohberg, and I. B. Spielman,
\textit{Light-induced gauge fields for ultracold atoms},
Rep. Prog. Phys. \textbf{77}, 126401 (2014).
\bibitem{goldman16}N. Goldman, J. C. Budich, and P. Zoller,
\textit{Topological quantum matter with ultracold gases in optical lattices},
Nat. Phys. \textbf{12}, 639 (2016).
\bibitem{sukumar97}C. V. Sukumar, D. M. Brink, \textit{Spin-flip transitions in
a magnetic trap}, Phys. Rev. A \textbf{56}, 2451 (1997).
\bibitem{burrows17}K. A. Burrows, H. Perrin, and B. M. Garraway,
\textit{Nonadiabatic losses from radio-frequency-dressed cold-atom traps:
Beyond the Landau-Zener model},
Phys. Rev. A \textbf{96}, 023429 (2017).
\bibitem{greschner13}
S. Greschner, L. Santos, and T. Vekua,
\textit{Ultra-cold bosons in zig-zag optical lattices},
Phys. Rev. A \textbf{87}, 033609 (2013).
\bibitem{dhar13}
A. Dhar, T. Mishra, R. V. Pai, S. Mukerjee, and B. P. Das,
\textit{Hard-core bosons in a zig-zag optical superlattice},
Phys. Rev. A \textbf{88}, 053625 (2013).
\bibitem{byrnes07}T. Byrnes, P. Recher, N. Y. Kim, S. Utsunomiya, and Y. Yamamoto,
\textit{Quantum Simulator for the Hubbard Model with Long-Range Coulomb
Interactions Using Surface Acoustic Waves}, Phys. Rev. Lett. \textbf{99}, 016405 (2007).
\bibitem{footnote2}
In principle, screening could also affect other system parameters, like, e.g.,
the effective electron mass in the 2DEG;
however, we do not expect this to play an important role since Coulomb drag effects
\cite{narozhny16}
between an electron and its image charge should be negligible due to the
high mobility of free electrons in the screening layer.
\bibitem{narozhny16}B. N. Narozhny, and A. Levchenko,
\textit{Coulomb drag},
Rev. Mod. Phys. \textbf{88}, 025003 (2016).
\bibitem{tosi14}G. Tosi, F. A. Mohiyaddin, H. Huebl, and A. Morello,
\textit{Circuit-quantum electrodynamics with direct magnetic coupling to single-atom spin qubits in isotopically enriched ${}^{28}$Si},
AIP Adv. \textbf{4}, 087122 (2014).
\bibitem{sarabi17}B. Sarabi, P. Huang, and N. M. Zimmerman,
\textit{Prospective two orders of magnitude enhancement in direct magnetic coupling of a single-atom spin to a circuit resonator},
arXiv:1702.02210 (2017).
\bibitem{ilin14}K. Ilin, D. Henrich, Y. Luck, Y. Liang, M. Siegel, and D. Yu. Vodolazov,
\textit{Critical current of Nb, NbN, and TaN thin-film bridges with and
without geometrical nonuniformities in a magnetic field},
Phys. Rev. B \textbf{89}, 184511 (2014).
%%%%%%%%%%%%%%%%%%%%%%%%%%%%%%%%%%%%%%%%%%%%%%%%%%%%%%%%%%%%%%%%%%%%%%%%
\bibitem{footnote1}
When the setup is operated in the high-frequency ($\omega \sim$ GHz) regime,
retardation effects of the propagating electromagnetic waves may become
important, which can be described by the
so-called \textit{Jefimenkov equations} \cite{jackson}.
However, corrections to the Biot-Savart law will be of the order of
$\sim d \omega / c$ where $c$ denotes the speed of light.
Even at ultra-high frequencies $\omega \sim (1 - 100)~$GHz and for
typical distances $d \sim (0.1-1)~\mu$m,
these corrections may safely be neglected and, in this quasistatic regime
($d \omega / c \ll 1$), the Biot-Savart law accurately describes the
induced magnetic field due to the current density $\mathbf{J}$.
%%%%%%%%%%%%%%%%%%%%%%%%%%%%%%%%%%%%%%%%%%%%%%%%%%%%%%%%%%%%%%%%%%%%%%%%
\bibitem{dreher12}L. Dreher, M. Weiler, M. Pernpeintner, H. Huebl, R. Gross, M. S. Brandt, and S. T. B. Goennenwein,
\textit{Surface acoustic wave driven ferromagnetic resonance in nickel thin films:
Theory and experiment}, Phys. Rev. B \textbf{86}, 134415 (2012).
\bibitem{landau35}L. Landau, E. Lifshitz.
\textit{On the Theory of the Dispersion of Magnetic Permeability in Ferromagnetic Bodies}.
Phys. Z. Sowjetunion \textbf{8}, 153 (1935).
\bibitem{gilbert04}T. Gilbert.
\textit{A phenomenological theory of damping in ferromagnetic materials}.
IEEE Trans. Magn. \textbf{40}, 3443 (2004).
\bibitem{rahav03}S. Rahav, I. Gilary, and S. Fishman,
\textit{Time Independent Description of Rapidly Oscillating Potentials},
Phys. Rev. Lett. \textbf{91}, 110404 (2003).
\bibitem{rahav03b}S. Rahav, I. Gilary, and S. Fishman,
\textit{Effective Hamiltonians for periodically driven systems},
Phys. Rev. A \textbf{68}, 013820 (2003).
\bibitem{sherman13}B. Sherman,
\textit{Optical generation of high amplitude laser generated surface acoustic waves},
AIP Conf. Proc. \textbf{1511}, 337 (2013).
\bibitem{schoen16}M. A. W. Schoen, D. Thonig, M. L. Schneider, T. J. Silva, H. T. Nembach, O. Eriksson, O. Karis, and J. M. Shaw,
\textit{Ultra-low magnetic damping of a metallic ferromagnet},
Nat. Phys. \textbf{12}, 839 (2016).
\bibitem{naber06}W. J. M. Naber, T. Fujisawa, H. W. Liu, and W. G. van der Wiel,
\textit{Surface-Acoustic-Wave-Induced Transport in a Double Quantum Dot},
Phys. Rev. Lett. \textbf{96}, 136807 (2006).
\bibitem{schuetz15}M. J. A. Schuetz, E. M. Kessler, G. Giedke, L. M. K. Vandersypen, M. D. Lukin, and J. I. Cirac,
\textit{Universal Quantum Transducers Based on Surface Acoustic Waves},
Phys. Rev. X \textbf{5}, 031031 (2015).
\bibitem{liu14}Y.-Y. Liu, K. D. Petersson, J. Stehlik, J. M. Taylor, and J. R. Petta,
\textit{Photon Emission from a Cavity-Coupled Double Quantum Dot},
Phys. Rev. Lett. \textbf{113}, 036801 (2014).
\bibitem{sousa03}R. de Sousa, and S. Das Sarmas,
\textit{Gate control of spin dynamics in III-V semiconductor quantum dots},
Phys. Rev. B \textbf{68}, 155330 (2003).
\bibitem{footnote3}Note that, in the literature, the Rabi frequency is
typically provided in the form $f_\text{Rabi}=\Omega_0/(2\pi) \approx \Omega_0 [\mu\mathrm{eV}] \times 240~$MHz.
\bibitem{kukushkin04}I. V. Kukushkin, J. H. Smet, L. H\"oppel, U. Waizmann,
M. Riek, W. Wegschneider, and K. von Klitzing,
\textit{Ultrahigh-frequency surface acoustic waves for finite wave-vector spectroscopy of two-dimensional electrons},
Appl. Phys. Lett. \textbf{85}, 4526 (2004).
\bibitem{koester96}S. J. Koester, B. Brar, C. R. Bolognesi, E. J. Caine, A. Patlach, E. L. Hu, H. Kroemer, and M. J. Rooks,
\textit{Length dependence of quantized conductance in ballistic constrictions fabricated on InAs/AlSb quantum wells},
Phys. Rev. B \textbf{53}, 13063  (1996).
\bibitem{yang02}C. H. Yang, M. J. Yang, K. A. Cheng, and J. C. Culbertson,
\textit{Characterization of one-dimensional quantum channels in InAs/AlSb},
Phys. Rev. B \textbf{66}, 115306  (2002).
\bibitem{koga11}T. Koga, S. Faniel, T. Matsuura, S. Mineshige, Y. Sekine, and H. Sugiyama,
\textit{Determination of Spin-Orbit Coefficients and Phase Coherence Times in InGaAs/InAlAs Quantum Wells}.
AIP Conf. Proc. \textbf{1416}, 38 (2011).
\bibitem{manchon15}A. Manchon, H. C. Koo, J. Nitta, S. M. Frolov, and R. A. Duine,
\textit{New perspectives for Rashba spin-orbit coupling}.
Nat. Mat. \textbf{14}, 871 (2015).
\bibitem{singleton01}J. Singleton,
\textit{Band Theory And Electronic Properties Of Solids},
Oxford University Press, New York (2001).
\bibitem{elzerman04}J. M. Elzerman, R. Hanson, L. H. Willems van Beveren,
B. Witkamp, L. M. K. Vandersypen, and L. P. Kouwenhoven,
\textit{Single-shot read-out of an individual electron spin in a quantum dot}.
Nature \textbf{430}, 431 (2004).
\bibitem{amasha08}S. Amasha, K. MacLean, I. P. Radu, D. M. Zumb\"uhl,
M. A. Kastner, M. P. Hanson, and A. C. Gossard,
\textit{Electrical Control of Spin Relaxation in a Quantum Dot}.
Phys. Rev. Lett. \textbf{100}, 046803 (2008).
\bibitem{khaetskii01}A. V. Khaetskii, and Y. V. Nazarov,
\textit{Spin-flip transitions between Zeeman sublevels in semiconductor quantum dots},
Phys. Rev. B \textbf{64}, 125316 (2001).
\bibitem{nadj-perge10}S. Nadj-Perge, S. M. Frolov, E. P. A. M. Bakkers, and L. P. Kouwenhoven,
\textit{Spin-orbit qubit in a semiconductor nanowire},
Nature \textbf{468}, 1084 (2010).
\bibitem{berg13}J. W. G. van den Berg, S. Nadj-Perge, V. S. Pribiag, S. R. Plissard, E. P. A. M. Bakkers, S. M. Frolov, and L. P. Kouwenhoven,
\textit{Fast Spin-Orbit Qubit in an Indium Antimonide Nanowire},
Phys. Rev. Lett. \textbf{110}, 066806 (2013).
\bibitem{gemelke05}N. Gemelke, E. Sarajlic, Y. Bidel, S. Hong, and S. Chu,
\textit{Parametric Amplification of Matter Waves in Periodically Translated Optical Lattices},
Phys. Rev. Lett. \textbf{95}, 170404 (2005).
\bibitem{lignier07}H. Lignier, C. Sias, D. Ciampini, Y. Singh, A. Zenesini, O. Morsch, and E. Arimondo,
\textit{Dynamical Control of Matter-Wave Tunneling in Periodic Potentials},
Phys. Rev. Lett. \textbf{99}, 220403 (2007).
\bibitem{struck11}J. Struck, C. \"Olschl\"ager, R. Le Targat, P. Soltan-Panahi, A. Eckardt, M. Lewenstein, P. Windpassinger, K. Sengstock,
\textit{Quantum Simulation of Frustrated Classical Magnetism in Triangular Optical Lattices},
Science \textbf{333}, 996 (2011).
\bibitem{struck12}J. Struck, C. \"Olschl\"ager, M. Weinberg, P. Hauke, J. Simonet, A. Eckardt, M. Lewenstein, K. Sengstock, and P. Windpassinger,
\textit{Tunable Gauge Potential for Neutral and Spinless Particles in Driven Optical Lattices},
Phys. Rev. Lett. \textbf{108}, 225304  (2012).
\bibitem{lacki16} M. \L{a}\c{c}ki, M. A. Baranov, H. Pichler, and P. Zoller,
\textit{Nanoscale Dark State Optical Potentials for Cold Atoms},
Phys. Rev. Lett. \textbf{117}, 233001 (2016).
\bibitem{wang18}Y. Wang, S. Subhankar, P. Bienias, M. La\c{c}ki, T.-C. Tsui, M. A. Baranov, A. V. Gorshkov, P. Zoller, J. V. Porto, and S. L. Rolston,
\textit{Dark State Optical Lattice with a Subwavelength Spatial Structure},
Phys. Rev. Lett. \textbf{120}, 083601 (2018).
\bibitem{movilla11}J. L. Movilla and J. Planelles,
\textit{Two-dimensional Bloch electrons under strong magnetic modulation},
Phys. Rev. B \textbf{83}, 014410 (2011).
\bibitem{bukov15}M. Bukov, L. D'Alessio, and A. Polkovnikov,
\textit{Universal high-frequency behavior of periodically driven systems:
from dynamical stabilization to Floquet engineering},
Adv. Phys. \textbf{64}, 139 (2015).
\bibitem{jackson}J. D. Jackson. \textit{Classical Electrodynamics}, Wiley, New York (1999).
\bibitem{gradshteyn00}
I. S. Gradshteyn and I. M. Ryzhik,
\textit{Table of Integrals, Series, and Products} (Academic Press, 2000), 6th ed.
\bibitem{weisstein}E. W. Weisstein,
\textit{Digamma function. From} MathWorld - \textit{A Wolfram Web Resource},
URL http://mathworld.wolfram. com/DigammaFunction.html.
\bibitem{robillard09}J.-F. Robillard, O. Bou Matar, J. O. Vasseur,
P. A. Deymier, M. Stippinger, A.-C. Hladky-Hennion, Y. Pennec, and B. Djafari-Rouhani,
\textit{Tunable magnetoelastic phononic crystals},
Appl. Phys. Lett. \textbf{95}, 124104 (2009).
\bibitem{leibfried03}D. Leibfried, R. Blatt, C. Monroe, and D. Wineland,
\textit{Quantum dynamics of single trapped ions},
Phys. Rev. Mod. \textbf{75}, 281 (2003).
\bibitem{trypogeorgos16}D. Trypogeorgos, and C. J. Foot,
\textit{Cotrapping different species in ion traps using multiple radio frequencies},
Phys. Rev. A \textbf{94}, 023609 (2016).
\bibitem{leefer17}N. Leefer, K. Krimmel, W. Bertsche, D. Budker, J. Fajans, R. Folman, H. H\"affner, F. Schmidt-Kaler,
\textit{Investigation of two-frequency Paul traps for antihydrogen production},
Hyperfine Int. \textbf{238}, 12 (2017).
\end{thebibliography}
\end{document}